\documentclass[journal]{IEEEtran}                                 

\IEEEoverridecommandlockouts     

\usepackage{comment}
\usepackage{booktabs} 

\usepackage{url}
\usepackage{diagrams}
\usepackage{color}
\usepackage{cite}
\usepackage{subfigure}

\usepackage{amssymb,latexsym,amsfonts,amsmath}
\renewcommand{\theequation}{\arabic{equation}}

\usepackage{tikz}
\usetikzlibrary{calc,shapes,arrows,intersections,automata,shapes,chains}
\usepackage[normalem]{ulem}
\usepackage{flushend}

\usepackage{enumitem}
 \setlist[enumerate,1]{label=\arabic*., leftmargin=0.5cm}
  \setlist[enumerate,2]{label=\alph*)., leftmargin=0.5cm}

\newcommand{\tup}{\textup}
\newcommand{\R}{{\mathbb{R}}}

\newcommand{\N}{{\mathbb{N}}}

\newcommand{\eg}{{\it e.g.}}

\newcommand{\KK}{\mathcal{K}_{\infty}}

\newcommand{\Let}{:=}

\definecolor{myco}{rgb}{0.55, 0.0, 0.63}

\newtheorem{theorem}{Theorem}[section]
\newtheorem{lemma}[theorem]{Lemma}
\newtheorem{proposition}[theorem]{Proposition}

\newtheorem{definition}[theorem]{Definition}
\newtheorem{example}[theorem]{Example}

\newtheorem{remark}[theorem]{Remark}

\def\bl#1{{\textcolor{black}{{\bf}#1}}}

\title{On Approximate Opacity of Stochastic Control Systems}

\author{Siyuan~Liu,~\IEEEmembership{Member,~IEEE,}
	Xiang~Yin,~\IEEEmembership{Member,~IEEE,}
       Dimos V. Dimarogonas,~\IEEEmembership{Fellow,~IEEE,} 
      and Majid~Zamani,~\IEEEmembership{Senior Member,~IEEE.}
        
\thanks{This work was supported in part by
the Horizon Europe EIC project SymAware (101070802), Digital Futures, the Knut and Alice Wallenberg (KAW) Foundation, the National Natural Science Foundation of China (62173226, 62061136004), 
the German Research Foundation (DFG) through the grant ZA 873/7-1, and the NSF under Grant ECCS-2015403. (Corresponding author: Xiang Yin.)}
	\thanks{S. Liu and D. Dimarogonas are with the Division of Decision and Control Systems,  KTH Royal Institute of Technology, Stockholm, Sweden. Email: {\tt\small \{siyliu,dimos\}@kth.se}}%
	\thanks{X. Yin is with the Department of Automation,  Shanghai Jiao Tong University and  Key Lab of System Control \& Information Processing,  Ministry of Education, Shanghai, China.
        Email: {\tt\small yinxiang@sjtu.edu.cn}}
\thanks{M. Zamani is with the Computer Science Department, University of Colorado Boulder, CO 80309, USA and the Computer Science Department, Ludwig Maximilian University of Munich, Germany.  Email: {\tt\small majid.zamani@colorado.edu}.   }     

}%

\begin{document}

\maketitle

\begin{abstract}
This paper investigates an important class of information-flow security property called opacity for stochastic control systems. Opacity captures whether a system's \emph{secret} behavior (a subset of the system's behavior that is considered to be critical) can be kept from outside observers. 
Existing works on opacity for control systems only provide a binary characterization of the system's security level by determining whether the system is opaque or not. 
In this work, we introduce a quantifiable measure of opacity that considers the likelihood of satisfying opacity for stochastic control systems modeled as general Markov decision processes (gMDPs). We also propose verification methods tailored to the new notions of opacity for finite gMDPs by using value iteration techniques.
Then, a new notion called \emph{approximate opacity-preserving stochastic simulation relation} is proposed, which captures the distance between two systems' behaviors in terms of preserving opacity. Based on this new system relation, we show that one can verify opacity for stochastic control systems using their abstractions (modeled as finite gMDPs). We also discuss how to construct such abstractions for a class of gMDPs under certain stability conditions. 
\end{abstract}


\vspace{-0.3cm}

\section{Introduction}
 
The analysis and synthesis of security for cyber-physical systems (CPS) have attracted considerable attentions over the past decades due to the increasing information exchange in modern CPS.
Various notions of security properties were considered in the literature for the protection of CPS against different types of attacks, which are typically classified into three categories along the CIA mnemonic: confidentiality, integrity, and availability~\cite{sandberg2015cyberphysical,dibaji2019systems,liu2022secure}.

This paper is concerned with a confidentiality property called \emph{opacity}.
In this problem, we assume that the system is monitored by an outside eavesdropper (malicious intruder) that is modeled as a passive observer. 
Opacity essentially characterizes the system's capability to conceal its \emph{secrets} information from being inferred by malicious intruders through the information flow \cite{saboori2007notions,lafortune2018history}.  
The concept of opacity is known to be originally introduced by \cite{mazare2004using} in the computer science literature to analyze security and privacy performances of cryptographic protocols. Since then, opacity has attracted extensive attentions in the discrete-event systems (DES) community.   
Depending on how the systems are modeled and what kind of secret the systems want to hide,   
different opacity notions have been proposed in the context of DES, including language-based notions \cite{lin2011opacity}, and state-based notions \cite{saboori2007notions,yang2020pre}.
Building on the various notions, verification and synthesis approaches were extensively developed for DES modeled as finite-state automata~\cite{saboori2012verification,yin2017new}, labeled transition systems~\cite{bryans2008opacity}, or Petri nets~\cite{tong2022verification}. The applications of opacity can be found in many real-world scenarios, including location-based services, mobile robot systems, and web services; see the survey papers   \cite{lafortune2018history,jacob2016overview} for more details on the historic developments of opacity. 

In more recent years, there has been a growing interest in the analysis of opacity for dynamical systems with continuous state-spaces \cite{liu2022secure,yin2020approximate,ramasubramanian2020notions,an2020opacity}. 
Particularly, different from the concepts of opacity appeared in DES literature, 
new notions of approximate opacity were proposed in \cite{yin2020approximate}, which are more applicable to continuous-space CPS whose outputs are physical signals. This new concept provides a quantitative evaluation of a system's security level by taking into account the imperfect measurements of external observations.
After that, various approaches have been proposed for the verification or synthesis of controllers with respect to approximate opacity for continuous-space control systems \cite{yin2020approximate,liu2020verification,mizoguchi2021abstraction,liu2021compositional}. 
In all the above-mentioned works, the definitions of opacity for dynamical systems only provide a binary characterization by simply reporting if the system is opaque or not. 
\bl{However, in many security-critical real-world applications, such as air traffic networks, infrastructure systems, robotic manufacturing, and biological systems, the systems inherently exhibit stochastic behavior due to the uncertainties and randomness in their operations and environments \cite{baier2008principles,lavaei2022automated}. Consequently, a binary characterization of whether a desired property is satisfied becomes inadequate and impractical. Given that the uncertainty affecting system evolution results in varying likelihoods of different system trajectories, it is more reasonable to evaluate the probability of a system being opaque. In fact, in many real-world scenarios, a small likelihood of violating opacity may be tolerable. Typical examples include the quantitative analysis of security in cloud computing systems \cite{zeng2019quantitative} and applications addressing location privacy issues \cite{mu2022verifying}, both of which rely on probabilistic quantification of opacity properties. Additionally, when a system is not secure in a non-stochastic sense, introducing a quantitative notion of opacity and developing corresponding probabilistic verification techniques allow us to assess the \emph{degree of satisfaction or violation} of the security property of interest. This approach can be instrumental in the early stages of system design.
}

\textbf{Our contribution.} 
In this paper, we propose a new concept of \emph{approximate opacity} 
for stochastic control systems modeled as general Markov decision processes (gMDPs) over continuous state spaces.
This new notion defines appropriate measures to quantify opacity in a probabilistic setting. 
We consider two basic types of opacity:  initial-state and current-state opacity.
In particular, a notion of \emph{approximate initial-state opacity} (resp. \emph{approximate current-state opacity}) is proposed for gMDPs which requires that any state run of the system should not violate initial-state opacity (resp. current-state opacity) with a probability greater than a threshold. 
Intuitively, the proposed new notions of opacity can be regarded as the probabilistic counterparts of the (qualitative) opacity notions in  \cite{yin2020approximate}, which allows us to quantitatively evaluate opacity for gMDPs. 
Based on these new notions of opacity, we further propose a verification approach for gMDPs using abstraction-based techniques.
In particular, new notions of opacity-preserving simulation relations are introduced to capture the distance between two gMDPs in terms of preserving approximate opacity. 
We show that one can verify approximate opacity of a (infinite) gMDP by using its finite abstractions (modeled as finite gMDPs) based on the these relations. 
In addition, we provide an approach to construct such finite abstractions for a class of discrete-time gMDPs. 
For the finite abstractions (finite gMDPs), we show that the verification of approximate opacity can be effectively achieved by using value iteration techniques. 


\textbf{Related works.} 
In the setting of stochastic DES, there have been several results in the literature that consider how to quantitatively evaluate opacity by using probabilistic models \cite{saboori2014current,berard2015probabilistic,chen2017quantification,ahmadi2018privacy,yin2019infinite}. 
For example, different probabilistic notions of current-state opacity were introduced in \cite{saboori2014current} for probabilistic finite-state automata; this approach has been extended to handle infinite-step and K-step opacity by \cite{yin2019infinite}. 
In \cite{chen2017quantification}, a Jensen-Shannon divergence-based measure was adopted to quantify secrecy loss in partially observed stochastic DES. 
Probabilistic measurement of opacity are also studied in  \cite{berard2015probabilistic,ahmadi2018privacy} for (partially-observed) Markov decision processes. 
Nevertheless, all the above-mentioned works are restricted to stochastic DES which evolves over discrete (finite) state spaces. 
In real-world applications, however, the gMDPs are equipped with continuous, and thus infinite, state sets. 
A suitably defined notion of opacity together with efficient verification techniques are indeed desirable in order to quantitatively evaluate the security level of general stochastic control systems.

There has been an attempt to extend the notion of opacity to stochastic control systems in \cite{liu2020stochastic}. 
Our results here differ from those in \cite{liu2020stochastic} in the following directions. 
First and foremost, the notion of opacity presented in \cite{liu2020stochastic} still provides a binary (qualitative) characterization of opacity, whereas the one proposed in the present paper is a quantitative probabilistic notion. 
Second, we develop abstraction-based verification approaches for both initial-state and current-state opacity here, whereas \cite{liu2020stochastic} only investigates initial-state opacity. Finally, effective verification algorithms are proposed here to verify different notions of opacity for finite gMDPs, which was not investigated in \cite{liu2020stochastic}.

\textbf{Organization.} 
The rest of the paper is organized as follows. In Section~\ref{sec:pre}, we introduce notations and necessary preliminaries on general Markov decision processes (gMDPs). In Section~\ref{sec:notion}, we propose new notions of approximate opacity for gMDPs. Verification approaches tailored to approximate opacity for finite gMDPs are presented in Section~\ref{sec:veri}. Section~\ref{Sec:systemrelations} introduces new notions of opacity-preserving stochastic simulation relations and discuss their properties. In Section~\ref{Sec:abs}, we describe how to construct finite abstractions for a class of stochastic control systems. A numerical case study is provided in Section \ref{sec:case} to illustrate the effectiveness of the proposed results. Finally, the paper is concluded in Section~\ref{sec:conclusion}.

\section{Preliminaries}\label{sec:pre}

\textbf{Notation.} 
 In this paper, a probability space is written as $(\Omega,\mathcal F_{\Omega},\mathbb{P}_{\Omega})$,
where $\Omega$ is the sample space,
$\mathcal F_{\Omega}$ is a sigma-algebra on $\Omega$ representing the set of events, and $\mathbb{P}_{\Omega}: \mathcal F_{\Omega} \rightarrow [0,1]$ is a probability measure that assigns probabilities to events.  
It is assumed that the random variables discussed in this paper are measurable functions of the form $X : (\Omega, \mathcal{F}_{\Omega}) \rightarrow (S_X, \mathcal{F}_X)$. Any random variable $X$ induces a probability measure on its measurable space $(S_X, \mathcal{F}_X)$ as $Prob\{A\} = \mathbb{P}_{\Omega}\{X^{-1}(A)\}$ for any $A \in \mathcal{F}_X$. For brevity, we often directly discuss the probability measure on $(S_X, \mathcal{F}_X)$ without explicitly mentioning the underlying probability space and the function $X$ itself. 
We further denote the set of all probability measures for a given measurable pair $(S_X, \mathcal{F}_X)$ as $\mathcal{P}(S_X, \mathcal{F}_X)$. 
We denote by $\R$ and $\N$ the set of real numbers and non-negative integers, respectively.
These symbols are annotated with subscripts to restrict them in
the usual way, \eg, $\R_{>0}$ denotes the positive real numbers. 
For $a,b\!\in\!\N$ and $a\!\le\! b$, we
use $[a;b]$ to
denote the corresponding intervals in $\N$.	
We use $\Vert x\Vert$ to denote the infinity norm of a vector $x \in \mathbb R^{n}$.
Given any $a\!\in\!\R$, $\vert a\vert$ denotes the absolute value of $a$. 
The composition of functions $f$ and $g$ is denoted by $f \circ g$.
We use notations $\mathcal{K}$ and $\mathcal{K}_{\infty}$ to denote the different classes of comparison functions, as follows: $\!\mathcal{K} \!\!=\! \!\{\gamma\!:\!\mathbb R_{\ge 0}\!\!\rightarrow\!\!\mathbb R_{\ge 0}  \! :  \! \gamma \text{ is continuous, strictly increasing and } \gamma(0)\!=\!0\}$; $\!\mathcal{K}_{\infty} \!\!=\! \!\{\gamma \!\in \!\mathcal{K} \! \!: \!  \lim\limits_{r\rightarrow \infty}\!\!\gamma(r) \!=\!\infty\}$.
For $\alpha$,$\gamma \!\in \!\mathcal{K}_{\infty}$ we write $\alpha\!<\!\gamma$ if $\alpha(s)\!<\!\gamma(s)$ for all $s\!>\!0$, and $\mathcal{I}_d\!\in\!\mathcal{K}_{\infty}$ denotes the identity function. 
Given sets $X$ and $Y$ with $X \! \subset  \! Y$, the complement of $X$ with respect to $Y$ is defined as $Y \setminus X \!=\! \{x \in Y:\!  x \!\notin\! X\}.$

\subsection{General Markov Decision Processes}
In this paper, we focus on \bl{a class of discrete-time stochastic control systems, which can be formally modeled as general Markov decision processes (gMDPs)} \cite{haesaert2017verification,lavaeilifting}.  
\begin{definition} \label{gMDP}
	A general Markov decision process (gMDP) is a tuple
	\begin{equation}
		\label{eq:dt-SCS}
		\Sigma\!=\!\left(X,U,T,Y, h\right)\!,
	\end{equation}
where $X\subseteq \mathbb R^n$, $U\subseteq \mathbb R^m$, and $Y\subseteq \mathbb R^q$ are Borel spaces denoting the state, external input, and output sets of the system, respectively. We use $\mathcal B(X)$ to denote the Borel sigma-algebra on the state set $X$, thus $(X, \mathcal B (X))$ denotes the corresponding measurable space. We use $T:\!\mathcal B(X) \!\times \!X \!\times\! U \!\rightarrow [0,1]$ to denote a conditional stochastic kernel that assigns to any state $x \in X$ and any input $u \in U$, a probability measure $T(\cdot|x,u)$ over the measurable space $(X, \mathcal B (X))$, so that for any set $\mathcal{A} \in \mathcal B (X)$,
	\begin{align}
		\mathbb{P}(x' \in \mathcal{A}|x,u) = \int_{\mathcal{A}} T(dx'|x,u).
	\end{align}
The measurable function $h:X\rightarrow Y$ is the output map.
\end{definition}
A system $\Sigma$ with a finite state set $X$ and finite input set $U$ is said to be a finite gMDP. 
The dynamical evolution of a gMDP can be equivalently described by 
\begin{equation}\label{dynamic}
	\Sigma:\left\{\hspace{-1.5mm}\begin{array}{l}\xi(k+1)=f(\xi(k),\nu(k),\varsigma(k)),\\
		\zeta(k)=h(\xi(k)),\\
	\end{array}\right.
	\quad k\in\mathbb N,
\end{equation} 
where $\xi(\cdot): \mathbb N\rightarrow X$, $\zeta(\cdot): \mathbb N\rightarrow Y$, and $\nu(\cdot):\mathbb N\rightarrow U$ are the state, output, and input signals, respectively. In the probability space $(\Omega,\mathcal F_{\Omega},\mathbb{P})$, we use $\varsigma  =  (\varsigma(0), \varsigma(1), \dots)$ to denote a sequence of independent and identically distributed (i.i.d.) random variables from $\Omega$ to the measurable set $V_\varsigma$, where $\varsigma(k) : \Omega \rightarrow  V_\varsigma,k  \in \mathbb{N}$.  
We use $\mathcal U$ to denote the collections of input sequences $\nu(\cdot):\mathbb N\rightarrow U$, where $\nu(k)$ is independent of $\varsigma(t)$ for any $k,t\in\mathbb N$ and $t\ge k$.  The map $f:X\times U\times V_\varsigma \rightarrow X$ is a measurable function serving as the state transition relation. 
Note that any gMDP endowed with the stochastic kernel $T$ as in Definition \ref{gMDP} can be equivalently represented by a gMDP as in \eqref{dynamic} with pair $(f, \varsigma)$ \cite{kallenberg1997foundations}. 
Throughout the paper, the system is assumed to be \emph{total}, i.e., for all $x \in X$, and for all $u \in U$, $f(x,u,\varsigma)$ is defined. 
\bl{
As described in Definition \ref{gMDP}, a gMDP operates over continuous or uncountable state sets, thereby extending an MDP by incorporating a metric output set where properties of interest are defined \cite{haesaert2017verification}. This modeling approach is sufficiently versatile to include specific cases such as discrete-time stochastic control systems \cite{lu}, labelled Markov processes (LMPs) \cite{desharnais2004metrics}, discrete-time stochastic hybrid games \cite{ding}, stochastic switched systems \cite{woong}, and other types of stochastic hybrid systems. 
}




Given any initial state $x_0$ and input sequence $\nu(\cdot) \in U$, the random sequences $ \xi_{x_0\nu}: \Omega \times \mathbb N \rightarrow X$ and $ \zeta_{x_0\nu}: \Omega \times \mathbb N \rightarrow Y$ satisfying \eqref{dynamic} are, respectively, called the \emph{state trajectory} and \emph{output trajectory} of $\Sigma$ under input sequence $\nu$ from $x_0$.  
Given an input sequence $\nu = (\nu_{0}, \nu_{1}, \dots, \nu_{L-1})$ over finite time horizon $L \in \mathbb{N}$, $\xi_{x_0\nu}= (x_0, x_1, \dots, x_L)$ is a \emph{finite state trajectory} of length $|\xi_{x_0\nu}| = L$ after initialization. 
For any  $\xi_{x_0\nu}=\! (x_0, x_1, \dots,  x_i, \dots, x_L)$, 
we use $\xi_{x_0\nu}(i) = x_i$ to denote the $i$th state of $\xi_{x_0\nu}$, and by  $\xi_{x_0\nu}[0;i] := (x_0, x_1, \dots, x_i)$ the  prefix of $\xi_{x_0\nu}$ with length $i$. 
Additionally, we drop the subscripts of $\xi_{x_0\nu}$ and denote by $\xi$ a state trajectory of the system when the initial state $x_0$ and input sequence $\nu$ are not of interest in some contexts. 
In this paper, we are interested in the state trajectories within a finite time horizon $n \in \mathbb{N}$. Thus, let us define $\mathcal{L}^n \Let \{ \xi:  |\xi| = n \}$ and $\mathcal{H}^n \Let \{ \zeta: |\zeta| = n \}$, respectively, as the set of all finite state trajectories and output trajectories of length $n \in \mathbb{N}$. We also denote by $\mathcal{U}^n := \{\nu(\cdot): [0; n-1] \rightarrow U \}$ the set of all input sequences over time horizon $n \in \mathbb{N}$.

\section{Approximate Opacity for gMDPs}\label{sec:notion}

In this section, we introduce new notions of (bounded-horizon)  approximate opacity quantifying the \emph{probability} of a gMDP preserving security within a given time horizon  $n \in \mathbb{N}$.   
Throughout this paper, we adopt a state-based formulation of secrets, i.e., the ``secrets" of the system as a set of states $X_S \subseteq X$.
Hereafter, we slightly modify the formulation in Definition \ref{gMDP} to accommodate for initial states and secret states, as $\Sigma = \left(X,X_0,X_S,U,T,Y, h\right)$, where $X_0 \subseteq X$ is a set of initial states and $X_S \subseteq X$ is a set of secret states.

Here we consider a passive \emph{intruder} (outside observer) who knows the dynamics of the system and can  observe the output sequences of the system without actively affecting the behavior of the system.
Furthermore, we assume that due to imperfect measurement precision, the intruder is unable to distinguish two ``$\varepsilon$-close" output trajectories $\zeta \!=\! (y_0, y_1, \dots, y_n)$ and $\zeta' \!=\! (y_0', y_1', \dots, y_n')$ if their distance is smaller than a given threshold $\varepsilon \geq 0$, i.e., $\max\limits_{i \in [0;n]}\Vert y_i \!-\!y_i' \Vert \leq \varepsilon$.
The two $\varepsilon$-close trajectories $\zeta$ and $\zeta'$ are said to be $\varepsilon$-approximate output-equivalent in the eye of an intruder with  measurement precision greater than $\varepsilon \geq 0$.  
Essentially, the intruder aims to infer certain secret information about the system using the output sequences observed online and the knowledge of the system model. Opacity essentially captures whether or not the system's secrets can be revealed to the intruder. 
For the sake of simplicity, we assume that the input sequences are internal information and unknown to the intruder. However, one could indeed drop this assumption to handle the setting where both input and output information are available to the intruder.

Before introducing the formal definitions of approximate opacity for gMDPs, let us define the sets of ``secure" state trajectories, i.e., the state trajectories whose occurrences do not reveal secrets to the passive intruder. 
Depending on what kind of information are regarded as critical, we consider two types of secrets, one w.r.t. initial-state information, and the other w.r.t. current-state information of the system. 


\textbf{Initial-State Secure Trajectories.}
For any state trajectory $\xi$ of length $|\xi| = L$, we say that $\xi$ \emph{does not reveal the initial secret} if there exists $\xi'$ of length $L$ such that  
\[
\xi'(0) \in X_0 \setminus\!  X_S\text{ and } \max\limits_{i \in [0;L]}\Vert h(\xi(i)) \!-\!h(\xi'(i)) \Vert \leq \varepsilon. 
\]
Essentially, the above indicates that for  $\xi$, there always exists an $\varepsilon$-approximate output-equivalent state trajectory initiated from a non-secret initial state.
A state trajectory $\xi$ is said to be \emph{initial-state secure} if it does not reveal the initial secret. We denote the set of initial-state secure state trajectories with  finite time horizon $n \in \mathbb{N}$ as:
\begin{align}  \label{secureinitialset}
\mathcal{L}_{\mathcal{I}_\varepsilon}^n\!\Let 
\!\left\{\xi\!\in\! \mathcal{L}^n: \xi \text{ is initial-state secure}\right\}. 
\end{align}


\textbf{Current-State Secure Trajectories.}
For any state trajectory $\xi$ of length $|\xi| = L$, we say that $\xi$ \emph{does not reveal the current secret} if there exists $\xi'$ of length $L$ such that
\[ \xi'(L) \in X \setminus\!  X_S\text{ and } \max\limits_{i \in [0;L]}\Vert h(\xi(i)) \!-\!h(\xi'(i)) \Vert \leq \varepsilon. 
\]
Similarly, the above says   that for  $\xi$, there always exists an $\varepsilon$-approximate output-equivalent state trajectory arriving at a non-secret current state. 
A state trajectory $\xi$ of length $L$ is said to be \emph{current-state secure} if  
none of its prefix $\xi[0;k], k \leq L$ reveals the current secret. 
We denote the set of  current-state  secure state trajectories with  finite time horizon $n \!\in \! \mathbb{N}$ as:
\begin{align}\label{securecurrentset}
    \mathcal{L}_{\mathcal{C}_\varepsilon}^n \Let \{\xi\!\in\! \mathcal{L}^n: \xi \text{ is current-state secure}\}.
\end{align}

Essentially,  upon observing any state trajectory in $\mathcal{L}_{\mathcal{I}_\varepsilon}^n$ (resp. $\mathcal{L}_{\mathcal{C}_\varepsilon}^n$), the intruder with imperfect measurement precision (captured by $\varepsilon$) is never certain that the initial (resp. current) state of the system is within the set of secret states. 
In this work, we are interested in establishing  \emph{a lower bound on the probability of the state trajectories of a gMDP being initial-state (or current-state) secure within a finite time horizon  $n \in \mathbb{N}$}. 
The formal definition of the new notions of approximate opacity for gMDPs is provided as follows.


\begin{definition}(\textbf{Approximate Opacity for gMDPs}) \label{opacity}
	Consider a gMDP $\Sigma\!=\! (X,X_0,X_S,U,T,Y,h)$ and constants $\varepsilon \!\in\! \R_{\geq 0}$, $0 \!\leq\! \lambda \!\leq\! 1$, and $n \!\in\! \mathbb{N}$. 
	System $\Sigma$ is said to be 
	\begin{enumerate}[leftmargin=*]
		\item[$\bullet$]  
		\emph{($\varepsilon, \lambda$)-approximate initial-state opaque} over time horizon $n$ if for any $x_0 \in X_0$ and any $\nu \in \mathcal{U}^n$, $\mathbb{P}_{x_0}^{\nu}(\xi_{x_0\nu} \in L_{\mathcal{I}_\varepsilon}^n) \geq \lambda$;
		\item[$\bullet$]  
		\emph{($\varepsilon, \lambda$)-approximate current-state opaque} over time horizon $n$ if  for any $x_0 \!\in\! X_0$ and any $\nu \!\in\! \mathcal{U}^n$, 
$\mathbb{P}_{x_0}^{\nu}(\xi_{x_0\nu} \in L_{\mathcal{C}_\varepsilon}^n) \geq \lambda$,
	\end{enumerate}
where $\mathbb{P}_{x_0}^{\nu}(\xi_{x_0\nu} \in L_{\mathcal{I}_\varepsilon}^n)$ (resp. $\mathbb{P}_{x_0}^{\nu}(\xi_{x_0\nu} \in L_{\mathcal{C}_\varepsilon}^n)$) denotes the probability that the state evolution of the system $\Sigma$ is initial-state (resp. current-state) secure over a finite time horizon $n$ starting from $x_0$ under input sequence $\nu$.
\end{definition}

Essentially, Definition~\ref{opacity} requires that, for any initial state and any input sequence, \bl{the system is probable to produce an initial-state/current-state secure trajectory  with a probability that is at least $\lambda$. The parameter $\varepsilon$ reflects the precision of measurements by potential external intruders, within which we can ensure the opacity of a gMDP.}
Without loss of generality,  throughout the work we assume  
\begin{equation}\label{initialassump}
    \forall x_0 \in X_0: \{x \in X_0: \Vert h(x_0)- h(x) \Vert \leq \varepsilon \} \nsubseteq X_S,
\end{equation}
for any gMDP $\Sigma=\left(X,X_0,X_S,U,T,Y, h\right)$;
otherwise ($\varepsilon, \lambda$)-approximate initial-state (resp. current-state) opacity is trivially violated.

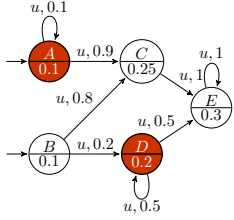
\begin{figure}[t!]
	\centering 
	{
		\resizebox{3.2cm}{!}{
			\begin{tikzpicture}[->,>=stealth',shorten >=1pt,auto,node distance=2.0cm,inner sep=1pt, initial text =,
				every state/.style={draw=black,fill=white,state/.style=state with output},
				accepting/.style={draw=black,thick,fill=red!80!green,text=white},bend angle=45]
				
				
				\node[state with output,accepting, initial] (A)   at (-1,0)                   {$A$ \nodepart{lower} $0.1$};
				\node[state with output]         (B) [right of=A] {$C$ \nodepart{lower} $0.25$};
				\node[state with output, initial] (C)   [below of=A]                 {$B$ \nodepart{lower} $0.1$};
				\node[state with output,accepting]         (D) [right of=C] {$D$ \nodepart{lower} $0.2$};
				\node[state with output]         (E) at (2.5, -1) {$E$ \nodepart{lower} $0.3$};

				\path (A) edge [loop above]    node {$u,0.1$} (A)				
				(A) edge              node {$u,0.9$} (B)
				
					(B) edge      node {$u,1$} (E)
				
				(C) edge    node {$u,0.8$} (B)
				edge           node {$u,0.2$} (D)

				(D) edge     [loop below]  node {$u,0.5$} (D)
				(D) edge         node {$u,0.5$} (E)
				
				(E) edge     [loop above]  node {$u,1$} (E)
				
				;
				
%
%
%
%
%

			\end{tikzpicture}
		}
	}
	\caption{States marked by red denote secret states, states marked by sourceless arrows denote initial states, and the output map is specified by the value associated to each state. For each transition, the input and state transition probability are also specified on the arrows.}
	\label{exautomata1}
	\vspace{-0.5cm}	
\end{figure}

We illustrate the proposed notions of approximate opacity for gMDPs using the following example.
\begin{example}
Consider a finite MDP  $\Sigma$ as shown in Fig.~\ref{exautomata1}, where $X= \{A, B,C,D,E\}$, $X_0 = \{A,B\}$, $X_S = \{A,D\}$, $U = \{u\}$, $Y=\{0.1,0.2,0.25,0.3\}$, and the output map is specified by the value associated with each state. The state transition probability equipped with each transition is also depicted on the arrows. First, we show that the system is $(\varepsilon, \lambda)$-approximate initial-state opaque within finite time horizon $n \in \mathbb{N}$ with $\varepsilon = 0, \lambda = 0.9$. 
By the definition of secure state trajectories in \eqref{secureinitialset}, 
we have for the secret initial state $A$, a finite state run in the form of $ACEE^*$ is initial-state secure, since there exists another state run starting from non-secret initial state $B$, i.e., $BCEE^*$ that generates exactly the same output run;
however, a finite state run in the form of $AAA^*CEE^*$ will generate an output run $(0.1,0.1,0.1^*,0.25,0.3,0.3^*)$, for which there does not exist a finite run starting from a nonsecret state that generates the same output run. 
Thus, we have for $x_0 = A$, $\mathbb{P}(\xi_{A\nu} \in L_{\mathcal{I}_0}^n) = 0.9 \geq \lambda$\footnote{In the remainder, we omit the sub- and subscripts of  $\mathbb{P}_{x_0}^{\nu}$ when they are clear from the context.}  for any $n \in \mathbb{N}$ and $\nu = u^*$. 
Notice that for the non-secret initial state $x_0 = B$, clearly we have $\mathbb{P}(\xi_{B\nu} \in L_{\mathcal{I}_0}^n) = 1 \geq \lambda$ for any $n \in \mathbb{N}$ and $\nu = u^*$. 
Therefore, the system is $(0, 0.9)$-approximate initial-state opaque within any finite time horizon $n  \in \mathbb{N}$. 
Similarly, one can check that the system is $(0, 0.8)$-approximate current-state opaque within any finite time horizon $n  \in \mathbb{N}$.
Note that this example is for illustrative purposes. We will provide a formal procedure for verifying approximate opacity for gMDPs in the next sections.
\end{example}

\begin{remark}
\bl{Let us remark that a notion of approximate opacity was proposed in \cite{yin2020approximate} for continuous-space control systems. The proposed notions in Definition \ref{opacity} can be seen as the probabilistic counterparts of the notions in \cite{yin2020approximate} to evaluate opacity for gMDPs quantitatively. Clearly, when analyzing non-stochastic control systems, Definition \ref{opacity} boils down to Definition III.1 in \cite{yin2020approximate} with $\lambda = 1$. Note also that several probabilistic opacity notions were proposed in [29] to provide a quantifiable measure of opacity. However, these notions are tailored to stochastic discrete event systems which can be modeled as probabilistic finite automata, and thus, not suitable for general stochastic control systems with continuous state spaces as considered in the present paper.
}    
\end{remark}

\begin{remark}
\bl{In simple terms, \emph{observability} \cite{kalman1960general} in control systems indicates that the system's states can be inferred from its external outputs, which is related to the concept of a system being non-opaque. The relationship between observability and opacity has been extensively explored in discrete-event systems (DES) literature. Specifically, \cite{lin2011opacity} demonstrated that observability in DES can be redefined as language-based opacity by appropriately defining the languages and observation mappings. Nevertheless, understanding the link between observability and opacity in general nonlinear (stochastic) control systems remains a complex task. This is due to the abundant notions of observability in the literature, such as \cite{kalman1960general,chen1980stochastic}, and is thus a subject for future research. }
\end{remark}

\begin{remark}
\bl{Note that the proposed notion of approximate opacity requires that for any initial state and any possible input sequence $\nu \in \mathcal{U}^n$, the probability that the resulting state trajectory is initial-state (or current-state) opaque must exceed a given threshold. Given this, the results presented in this paper are not limited by any specific control policy imposed on the system. Therefore, the proposed results can be directly applied to stochastic control systems under any pre-designed control policy by restricting the set of possible control sequences, such as the discrete feedback or sampled-data feedback controllers developed in existing literature \cite{chen2021mean,chen2023controller,chen2021sampled}. It would also be an interesting future research direction to explore the design of discrete or sampled-data feedback controllers to enforce opacity for gMDPs using approaches similar to those in these references.} 
\end{remark}

\vspace{-0.3cm}

\section{Verification of Approximate Opacity for Finite GMDPs}\label{sec:veri}
In this section, we show how to verify approximate opacity for finite gMDPs. The results here will provide a foundation for the verification of approximate opacity for infinite gMDPs through certain system relations as shown in Section \ref{Sec:systemrelations}.

\subsection{Verification of Approximate Initial-State Opacity}\label{subsec:veriinitial}
In order to verify approximate initial-state opacity for finite gMDPs, we will first construct a new system which serves as an approximate initial-state estimator. For the sake of brevity, for any $x\in X$,  we define 
\[
B_\varepsilon(x)=\{ x'\in X: \Vert h(x)\!-\!h(x') \Vert \leq\! \varepsilon\}
\]
as the set of states that are $\varepsilon$-close to $x$ in terms of output observations. 
The formal definition of an approximate initial-state estimator is provided as follows.
\begin{definition}\label{def:iniesti}
(\textbf{Approximate initial-state estimator})
Consider a finite gMDP  $\Sigma\!=\!(X,X_0,X_S,U,T,Y, h)$ and  $\varepsilon \in \R_{\geq 0}$. 
The $\varepsilon$-approximate initial-state estimator is a finite gMDP (without outputs) 
$\Sigma_{I,obs}=(X_I, X_{I0}, U,  T_I)$,
where 
\begin{itemize}
	\item 
	$X_I=\{ (x,q)\in X\times   2^{X\times X}: (\bar{x}_1,\bar{x}_2)\in q \Rightarrow \bar{x}_2 \in B_\varepsilon(x)\}$ is the set of states; 
	\item 
	$X_{I0}=\{ (x,q)\in X_0\times 2^{X_0\times X_0}:  (\bar{x}_1,\bar{x}_2)\!\in\! q \Leftrightarrow [\bar{x}_1\!=\!\bar{x}_2 \wedge \bar{x}_2 \in B_\varepsilon(x)]\}$ is the set of initial states; 
	\item $U$ is the same set of inputs as in $\Sigma$;   
	\item 
	$T _I: X_I\times U\times  X_I \to [0,1]$ is the transition probability function defined by: for any $x_I=(x,q)\in X_I$, any $x_I'=(x',q')\in X_I$, and any $u\in U$, 
	\begin{itemize}
		\item 
		$T_I(x_I' | x_I, u)\neq 0$ if 
		\begin{itemize}
			\item  
			$ T(x'|x, u)> 0$, and 
			\item
			$q'= \bigcup_{{u'}\in U} \{(\bar{x}_1,\bar{x}_2')\in X\times X: (\bar{x}_1,\bar{x}_2)\in q\wedge T(\bar{x}_2'|\bar{x}_2, u')> 0 \wedge \bar{x}_2'\in B_{\varepsilon}(x') \}   $.
		\end{itemize}
		\item 
		when $T_I(x_I' | x_I, u)\neq 0$, we have 
		$T_I(x_I' | x_I, u) = T(x'| x, u)   $.
	\end{itemize}
\end{itemize}
\end{definition}
In the above definition, for each $(x,q) \in X_I$, $x$ is the real system state, and $q$ indicates the initial-state estimate. 
Specifically, $q\in 2^{X\times X}$ is a set of state pairs, i.e., each state in $q$ is in the form of $(\bar{x}_1,\bar{x}_2)$, where $\bar{x}_1$ denotes the  state where the trajectory comes from initially and $\bar{x}_2$ denotes a possible current state that is $\varepsilon$-close to its real current state.  
Thus, the first component of $q$ is the initial-state estimate. 
Note that from the construction of $\Sigma_{I,obs}$, for any state trajectory $(x_0,\dots,x_n)$ of $\Sigma$, there exists a unique initial-state estimation sequence $((x_0,q_0)\dots(x_n,q_n))$ in its approximate initial-state estimator $\Sigma_{I,obs}$. 
In addition, we denote by $\xi^I_{x_0}$ a state trajectory of  $\Sigma_{I,obs}$ generated from initial state $(x_0,q_0) \in X_{I0}$ under some input sequence.

Given any state trajectory $(x_0,\dots,x_n)$ of $\Sigma$ and its output sequence $(h(x_0),\dots,h(x_n))$, let us define the initial-state estimate with $\varepsilon$ precision by observing $(h(x_0),\dots,h(x_n))$ as  
\begin{align} 
&E_I^\varepsilon(h(x_0),\dots,h(x_n)) := \label{estimate_ini}\\
&\{  \bar x_0\!\in\! X_0: \exists (\bar x_0,\dots, \bar x_n)\!\in\! \mathcal{L}^n  
 \text{ s.t. }\max_{i \in [0;n]} \Vert h(\bar x_i)\!- \!h(x_i) \Vert \!\leq \! \varepsilon \}.  \notag
\end{align}
Consider the corresponding initial-state estimation sequence $((x_0,q_0)\dots(x_n,q_n))$ in $\Sigma_{I,obs}$.
Since the first component of $q_n$ in $\Sigma_{I,obs}$ is the initial-state estimate, we have $E_I^\varepsilon(h(x_0),\dots,h(x_n)) =\{\bar{x}_1 \in X: (\bar{x}_1,\bar{x}_2)\in q_n\}.$ 

Accordingly, we can define the set of ``bad" (secret revealing) states in $\Sigma_{I,obs}$ as:
\begin{align}\label{badset}
\!\!\!\!X_I^{bad}:=\{ (x,q)\in X_I:  \{\bar{x}_1\in X: (\bar{x}_1,\bar{x}_2)\in q\} \subseteq X_S \}.
 \end{align}
Here, a ``bad" state is a state in the initial-state estimator whose associated initial-state estimate lies entirely within the set of secret states.

Note that by fixing  $\nu(t)\in U$ in $\Sigma_{I,obs}$ at any time step $t \geq 0$, we obtain a transition probability matrix denoted by 
$\mathbf{P}_I^{\nu(t)}: X_I\times X_I \to [0,1]$. 
Then the next two lemmas are needed to prove the main result in this subsection.
\begin{lemma}\label{lemma:1_ini}
Consider a finite gMDP  $\Sigma\!=\!(X,X_0,X_S,U,T,$ $Y,h)$ and its approximate initial-state estimator $\Sigma_{I,obs}$ as in Definition~\ref{def:iniesti}.
For any initial state $x_0$ and any input sequence  $\nu= (\nu(0), \dots, \nu(n-1))$, we have 
\[
1-\mathbb{P}(\xi_{x_0\nu} \in L_{\mathcal{I}_\varepsilon}^n)
= \mathbf{1}^I_{bad} \mathbf{P}_I^{\nu(n-1)}\cdots  \mathbf{P}_I^{\nu(1)} \mathbf{P}_I^{\nu(0)} \mathbf{1}^I_{x_0},
\]
where $L_{\mathcal{I}_\varepsilon}^n$ is the set of initial-state secure state trajectories as in \eqref{secureinitialset} within time horizon $n \in \mathbb{N}$, and $\mathbf{1}^I_{bad}$ and $\mathbf{1}^I_{x_0}$ are the indicator vectors\footnote{Specifically, $\mathbf{1}^I_{bad}$ is a binary row vector indexed by states in $X_I$ such that each element whose index is in $X_I^{bad}$ is $1$, otherwise it is $0$. Similarly, given any $x_0$, $\mathbf{1}^I_{x_0}$ is a binary row vector indexed by states in $X_I$ such that each element with index $(x_0,q_0) \in X_{I0}$ (with the first component being $x_0$) is $1$, otherwise it is $0$. Note that $\mathbf{1}^I_{x_0}$  are \emph{standard basis vectors} for any $x_0 \in X_0$.} for states in $X_I^{bad}$ as defined in \eqref{badset} and for initial states in $X_{I0}$ in $\Sigma_{I,obs}$, respectively. 
\end{lemma}

The proof of Lemma \ref{lemma:1_ini} is provided in the Appendix.



\begin{lemma}\label{lemma:opa_ini}
Consider a finite gMDP  $\Sigma=(X,X_0,X_S,U,T,$ $Y,h)$ and its approximate initial-state estimator $\Sigma_{I,obs}$ as in Definition~\ref{def:iniesti}.
If for any initial state $x_0$ and any input sequence  $\nu\!=\! (\nu(0),  \dots, \nu(n-1))$, the following holds: 
\begin{align}\label{verifycondition_ini}
\mathbf{1}^I_{bad} \mathbf{P}_I^{\nu(n-1)}\cdots  \mathbf{P}_I^{\nu(1)} \mathbf{P}_I^{\nu(0)} \mathbf{1}^I_{x_0}
  \leq  1-\lambda,
\end{align}
where $\mathbf{1}^I_{bad}$ and $\mathbf{1}^I_{x_0}$ are the indicator vectors for the states in $\Sigma_{I,obs}$  as in Lemma~\ref{lemma:1_ini}, then $\Sigma$ is $(\varepsilon, \lambda)$-approximate initial-state opaque over time horizon $n \in \mathbb{N}$.
\end{lemma}

The proof of Lemma \ref{lemma:opa_ini} is provided in the Appendix.
Next, we show how to verify approximate initial-state opacity for finite gMDPs using value iteration techniques.

\textbf{Bounded-Step Reachability. }
For a finite gMDP  $\Sigma_{I,obs}$ as in Definition~\ref{def:iniesti}, the problem of checking whether \eqref{verifycondition_ini} holds can be solved by computing the \emph{maximal probability of reaching the ``bad" states in $X_I^{bad}$, as defined in \eqref{badset}, within finite time horizon $n \in \mathbb{N}$  starting from any initial state $(x_0,q_0) \in X_{I0}$ under any input sequence $\nu$}. 
This \emph{bounded-step reachability} property is denoted by $\lozenge^{\leq n} X_I^{bad}$, where $\lozenge$ denotes the ``eventually" operator. 
Let us define the maximal probabilities of the above mentioned reachability problem as:
\begin{equation} \label{maximalP_ini}
    \mathbb{P}^{\max}(\xi^I_{x_0}\models \lozenge^{\leq n}  X_I^{bad}),
\end{equation}
where $\xi^I_{x_0}\models \lozenge^{\leq n} X_I^{bad}$ represents that the state trajectory $\xi^I_{x_0}$ of $\Sigma_{I,obs}$ satisfies the property of ``eventually reaching set $X_I^{bad}$ within $n \in \mathbb{N}$ time steps".  
Let us further define the vector $\left(p_{s}\right)_{s \in X_{I}}$, with $p_{s}:= \mathbb{P}^{\max}(\xi^I_{s}\models \lozenge^{\leq n} X_I^{bad})$ being the maximal probability of reaching $X_I^{bad}$ within $n \in \mathbb{N}$ time steps starting from state $s \in X_{I}$ under any input sequence.
We further denote by $X_I^{nb} \subseteq X_I \setminus X_I^{bad}$  the set of states from which $X_I^{bad}$ is not reachable in the underlying directed graph of the finite gMDP $\Sigma_{I,obs}$.

We show in the next lemma that the above-discussed maximal reachability probability of the finite gMDP  $\Sigma_{I,obs}$ can be computed by a value iteration approach. This lemma is inspired by \cite[Thm.10.100]{baier2008principles} and provided without proof.

\begin{lemma}\label{lemma:LP}
(Value iteration for bounded-step reachability probabilities of finite gMDPs)
Consider a finite gMDP $\Sigma_{I,obs}=(X_I, X_{I0}, U,  T_I)$ as in Definition~\ref{def:iniesti} and set  $X_I^{bad} \subseteq X_I$ as defined in \eqref{badset}. 
The vector $\left(p_{s}\right)_{s \in X_{I}}$ with  $p_{s}= \mathbb{P}^{\max}(\xi^I_{s}\models \lozenge^{\leq n} X_I^{bad})$ can be computed by following the value iteration process:
  \begin{align*}
 & \textbf{1)} \ p_s = p_s^{(i)} = 1,  \text{ for any } i, \text{ for all }  s \in X_I^{bad}; \\
	& \textbf{2)} \ p_s = p_s^{(i)} = 0,  \text{ for any } i, \text{ for all }  s \in X_I^{nb};\\
	& \textbf{3)} \  p_s =  p_s^{(n)},  \text{ for all }  s \in X_I \setminus (X_I^{bad} \cup X_I^{nb}), \\
& \quad  \text{ with }   p_s^{(0)} = 0, p_s^{(i+1)} = \max \{\sum_{s' \in X_{I}}T_I(s'| s, u) p_s'^{(i)}\}.  
\end{align*}
\end{lemma}

By using the value iteration approach in the above lemma, one can get the maximal probabilities $\mathbb{P}^{\max}(\xi^I_{x_0}\models \lozenge^{\leq n} X_I^{bad})$ for all initial states $x_0 \in X_0$.
We discuss the complexity of computing $\mathbb{P}^{\max}(\xi^I_{s}\models \lozenge^{\leq n} X_I^{bad})$ later in Remark \ref{remark:computation}.

In the next theorem, we present the main result in this subsection on the verification of approximate initial-state opacity of finite gMDPs.
\begin{theorem}\label{thm:veri_ini}
Consider a finite gMDP  $\Sigma\!=\!(X,X_0,X_S,U,$ $T,Y,h)$ and its approximate initial-state estimator $\Sigma_{I,obs}$ as in Definition~\ref{def:iniesti}.
Then, $\Sigma$ is $(\varepsilon, \lambda)$-approximate initial-state opaque over time horizon $n \!\in\! \mathbb{N}$ if and only if for any $x_0 \in X_0$:
$$
\mathbb{P}^{\max}(\xi^I_{x_0}\models \lozenge^{\leq n} X_I^{bad}) \leq 1-\lambda,
$$
with $\mathbb{P}^{\max}(\xi^I_{x_0}\models \lozenge^{\leq n} X_I^{bad})$ computed as in Lemma \ref{lemma:LP}.
\end{theorem}
\begin{IEEEproof}
First recall from \eqref{maximalP_ini} that $\mathbb{P}^{\max}(\xi^I_{x_0}\models \lozenge^{\leq n} X_I^{bad})$ denotes the maximal probability of a trajectory $\xi^I_{x_0}$ in $\Sigma_{I,obs}$ eventually reaching $X_I^{bad}$ within time horizon $n \in \mathbb{N}$ under any input sequence. Thus, we have for any $x_0 \in X_0$
$$\mathbb{P}^{\max}(\xi^I_{x_0}\models \lozenge^{\leq n} X_I^{bad}) \! =\! \max_{\nu} \mathbf{1}^I_{bad} \mathbf{P}_I^{\nu(n-1)}\!\cdots  \mathbf{P}_I^{\nu(1)} \mathbf{P}_I^{\nu(0)} \mathbf{1}^I_{x_0}$$
where $\!\nu\!=\! (\nu(0), \!\dots,\! \nu(n\!-\!1))$.
Then, \eqref{verifycondition_ini} holds if and only if $\mathbb{P}^{\max}(\xi^I_{x_0}\models \lozenge^{\leq n} X_I^{bad}) \!\leq\! 1\!-\!\lambda$.  
The rest of the proof for this theorem follows readily by combining the results in Lemmas \ref{lemma:1_ini} and \ref{lemma:opa_ini}.
\end{IEEEproof}

\begin{example}\label{example:veri}	
\begin{figure}
		\centering
		\includegraphics[width=0.455\textwidth]{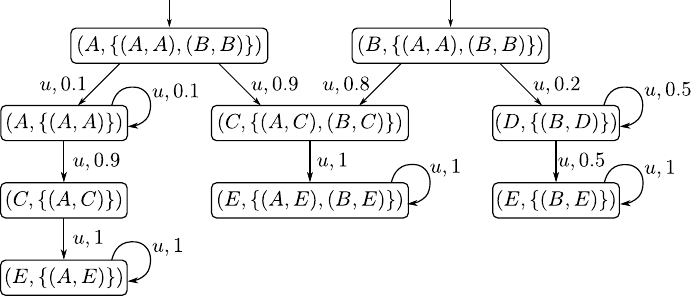}
		\caption{Example of approximate initial-state estimators.} 
		\label{fig:initial_estimator}
  \vspace{-0.5cm}
\end{figure}
Let us consider the finite gMDP $\Sigma$ as shown in Fig.~\ref{exautomata1}. In order to verify approximate opacity of $\Sigma$, we first build an $\varepsilon$-approximate initial-state estimator $\Sigma_{I,obs}$ following Definition \ref{def:iniesti}, with $\varepsilon = 0$. 
The estimator is shown in Fig.~\ref{fig:initial_estimator}, where 
$X_I  = \{(A,$ $ \{(A,A), (B,B)\}), (B, \{(A,A), (B,B)\}), (A, \{(A,A)\}), (C, $ $\{(A,C), (B,C)\}), (D, \{(B,D)\}), (C, \{(A,C)\}), (E, \{(A,E),$ $ (B,E)\}), (E, \{(B,E)\}), (E, \{(A,E)\})\}$, $U = \{u\}$, $X_{I0} = $ $\{ (A,\{(A,A),(B,B)\}),(B,\{(A,A),(B,B)\})\}$.
By \eqref{badset}, we get $X_I^{bad} = $ $\{(A, \{(A,A)\}), (C, \{(A,C)\}), (E, \{(A,E)\})\}$, from which one can readily have the indicator vectors $\mathbf{1}^I_{bad} = \left[ 0, 0, 1, 0, 0 , 1 , 0 ,  0 , 1 \right]$, $\mathbf{1}^I_{A} = \left[ 1, 0, 0, 0, 0 , 0 , 0 ,  0 , 0 \right]$ and $\mathbf{1}^I_{B} =\left[ 0, 1, 0, 0, 0 , 0 , 0 ,  0 , 0 \right]$. 
Now, considering $\lambda = 0.9$, we show that $\Sigma$ is (0,0.9)-approximate initial-state opaque by applying Theorem \ref{thm:veri_ini}.
Let us first compute  $\mathbb{P}^{\max}(\xi^I_{x_0}\models \lozenge^{\leq n} X_I^{bad})$ by following the value iteration process as in Lemma \ref{lemma:LP}. 
Note that based on the underlying directed graph of this finite gMDP, we have $X_I^{nb} = \{(B, \{(A,A), (B,B)\}),  (C, \{(A,C), (B,C)\}), (D, \{(B,D)\}),$ $(E, \{(A,E), (B,E)\}), (E, \{(B,E)\})\} $ as the set of states from which $X_I^{bad}$ is not reachable. 
The equations of the value iteration process are: 
1)
for the states in $X_I^{bad}$, we get $p_{(A, \{(A,A)\})}^{(i)} = p_{(C, \{(A,C)\})}^{(i)} = p_{(E, \{(A,E)\})}^{(i)} = 1$ for any $i$; 2) for the states in $X_I^{nb}$, we get $p_{(B, \{(A,A), (B,B)\})}^{(i)} = p_{(C,\{(A,C), (B,C)\})}^{(i)} = p_{(D, \{(B,D)\})}^{(i)} = p_{ (E, \{(A,E),  (B,E)\})}^{(i)} = p_{(E, \{(B,E)\})}^{(i)} = 0$  for any $i$; 
3) for the other state $(A,\{(A,A),(B,B)\})$ which is neither in $X_I^{bad}$ nor in $X_I^{nb}$, we have $p_{(A,\{(A,A),(B,B)\})}^{(0)} = 0$, 
$p_{(A,\{(A,A),(B,B)\})}^{(i+1)} = \max\{0.1p_{(A,\{(A,A),(B,B)\})}^{(i)}, 0.9p_{(C,\{(A,C), (B,C)\})}^{(i)}\} = 0.1$. Thus, the solution of the value iteration process  in Lemma \ref{lemma:LP} yields the vector $\left(p_{s}\right)_{s \in X_{I}} = \{0.1, 1,1,1,0, 0, 0, 0, 0\}$ with $n \geq 1$. 
Therefore, we obtain that $\mathbb{P}^{\max}(\xi^I_{x_0}\models \lozenge^{\leq n} X_I^{bad}) = 0.1$ for $x_0 = A$ and $\mathbb{P}^{\max}(\xi^I_{x_0}\models \lozenge^{\leq n} X_I^{bad}) = 0$ for $x_0 = B$, which shows that $\mathbb{P}^{\max}(\xi^I_{x_0}\models \lozenge^{\leq n} X_I^{bad}) \leq 1-\lambda$ for any initial state $x_0$.  
By Theorem \ref{thm:veri_ini}, we conclude that $\Sigma$ is (0,0.9)-approximate initial-state opaque within any time horizon $n \in \mathbb{N}$ with $n \geq 1$.
\end{example}

\vspace{-0.2cm}
\subsection{Verification of Approximate Current-State Opacity}
 
In this subsection, we show how to verify approximate current-state opacity for finite gMDPs. To achieve this, we first construct a new system that serves as an approximate current-state estimator as follows.

\begin{definition}\label{estimator:current}	
(\textbf{Approximate current-state estimator})
	Consider a finite gMDP  $\Sigma\!=\!(X,X_0,X_S,U,T,Y, h)$ and  $\varepsilon \in \R_{\geq 0}$. 
	The $\varepsilon$-approximate current-state estimator is a finite gMDP (without outputs) $\Sigma_{C,obs}=(X_C, X_{C0}, U, T_I)$, where 
	\begin{itemize}
		\item 
		$X_C=\{ (x,q,l)\in X\times   2^{X} \times \{0,1\}: \bar{x}\in q \Rightarrow \bar{x} \in B_\varepsilon(x)\}$ is the set of states; 
		\item 
		$X_{C0}=\{ (x,q,l)\in X_0\times 2^{X_0}\times \{0,1\}:  [\bar{x}\in q \Leftrightarrow \bar{x}\in B_\varepsilon(x)] \wedge [l = 0]\}$ is the set of initial states;
		\item $U$ is the same set of inputs as in $\Sigma$;   
		\item 
		$T_C: X_C\times U\times  X_C \to [0,1]$ is the transition probability function defined by: for any $x_C=(x,q,l)\in X_C$, any $x_C'=(x',q',l')\in X_C$, and any $u\in U$, 
		\begin{itemize}
			\item 
			$T_C(x_C' | x_C, u)\neq 0$ if 
			\begin{itemize}
				\item  
				$ T(x'|x, u)> 0$, 
				\item
				$q'= \bigcup_{u'\in U} \{ \bar{x}'\in X:  \bar{x} \in q \wedge T(\bar{x}'|\bar{x}, u')> 0 \wedge \bar{x}'\in B_{\varepsilon}(\bar{x}') \}   $, and
    \item $l' = 0 \Leftrightarrow [l =0] \wedge [\{\bar{x}'\in X: \bar{x}'\in q'\} \nsubseteq X_S]$.
			\end{itemize}
			\item 
			when $T_C(x_C' | x_C, u)\neq 0$, we have 
			$T_C(x_C' | x_C, u) = T(x'| x, u)   $.
		\end{itemize}
	\end{itemize}
 
\end{definition}

In the above definition, for each $(x,q,l) \in X_C$, $x$ is the real system state, $q$ is the current-state estimate, and $l$ is a binary state taking values either $0$ or $1$. 
Note that $q \in 2^X$ is a set of states, with $\bar x \in q$ denoting its current state estimate. 
According to Definition \ref{estimator:current}, for a state $(x',q',l') \in X_C$, $l'=0$ if and only if: 1) $\{\bar{x}'\in X: \bar{x}'\in q'\} \nsubseteq X_S$, i.e., the current-state estimate does not lie entirely within the set of secret states, and 2) $l=0$ at the previous time step. Otherwise, $l'=1$. 
Essentially, for any trajectory that is currently at state $(x,q,l)$, 
$l$ tracks whether or not the trajectory has revealed the current-state secret before or upon visiting the state. In particular, $l=0$ indicates that none of the prefixes of the trajectory reveals the current-state secret; 
$l=1$ indicates that the trajectory reveals the current-state secret at the current step or previous time steps.

For any state trajectory $(x_0,\dots,x_n)$ of $\Sigma$, there exists a unique current-state estimation sequence $((x_0,q_0,l_0)\dots(x_n,q_n,l_n))$ in its approximate current-state estimator $\Sigma_{C,obs}$. 
In addition, we denote by $\xi^C_{x_0}$ a state trajectory of  $\Sigma_{C,obs}$ generated from initial state $(x_0,q_0,l_0) \in X_{C0}$ under some input sequence.
Accordingly, we can define the set of ``bad" (secret revealing) states as:
\begin{equation}\label{badset:cur}
 X_C^{bad}=\{ (x,q,l)\in X_C:  l =1  \}.
\end{equation}


By fixing an input $\nu(t)\in U$ in $\Sigma_{C,obs}$ at each time step $t \geq 0$, we obtain a transition probability matrix denoted by 
$\mathbf{P}_C^{\nu(t)}: X_C\times X_C \to [0,1]$.  
Similar as stated in Lemma~\ref{lemma:1_ini} and Lemma~\ref{lemma:opa_ini}, given any finite gMDP  $\Sigma\!=\!(X,X_0,X_S,U,T,$ $Y,h)$ and its approximate current-state estimator $\Sigma_{C,obs}$ as in Definition~\ref{estimator:current}, for any initial state $x_0$ and any input sequence  $\nu= (\nu(0), \dots, \nu(n-1))$, we have 
\[
1-\mathbb{P}(\xi_{x_0\nu} \in L_{\mathcal{C}_\varepsilon}^n)
=  \mathbf{1}^C_{bad} \mathbf{P}_C^{\nu(n-1)}\cdots  \mathbf{P}_C^{\nu(1)} \mathbf{P}_C^{\nu(0)} \mathbf{1}^C_{x_0},
\]
where $L_{\mathcal{C}_\varepsilon}^n$ is the set of current-state secure state trajectories as in \eqref{securecurrentset} within time horizon $n \in \mathbb{N}$, and $\mathbf{1}^C_{bad}$ and $\mathbf{1}^C_{x_0}$ are the indicator vectors
for states in $X_C^{bad}$  as defined in \eqref{badset:cur} and for initial states in $X_{C0}$ in $\Sigma_{C,obs}$, respectively.
Moreover, if any initial state $x_0$ and any input sequence  $\nu= (\nu(0), \dots, \nu(n-1))$, the following holds: 
\begin{align}\label{verifycondition_cur}
\mathbf{1}^C_{bad} \mathbf{P}_C^{\nu(n-1)}\cdots  \mathbf{P}_C^{\nu(1)} \mathbf{P}_C^{\nu(0)}   \mathbf{1}^C_{x_0} \leq 1-\lambda,
\end{align}
then $\Sigma$ is $(\varepsilon, \lambda)$-approximate current-state opaque  over time horizon $n \in \mathbb{N}$.
The verification of approximate current-state opacity for finite gMDPs follows similarly as in Subsection \ref{subsec:veriinitial} using value iteration techniques.
Let us denote by  $\mathbb{P}^{\max}(\xi^C_{x_0}\models \lozenge^{\leq n} X_C^{bad})$ 
the maximal probabilities of a trajectory $\xi^C_{x_0}$ in $\Sigma_{C,obs}$ reaching $X_C^{bad}$ within bounded time horizon $n \in \mathbb{N}$ under any input sequences. 
Note that by following the value iteration approach similar to Lemma \ref{lemma:LP}, one can compute $\mathbb{P}^{\max}(\xi^C_{s}\models \lozenge^{\leq n} X_C^{bad})$ as well. 



The next theorem summarizes the main result of the verification of approximate current-state opacity of finite gMDPs.
\begin{theorem}\label{thm:veri_cur}
Consider a finite gMDP  $\Sigma\!=\!(X,X_0,X_S,U,$ $T,Y,h)$ and its approximate current-state estimator $\Sigma_{C,obs}$ as in Definition~\ref{estimator:current}.
Then, $\Sigma$ is $(\varepsilon, \lambda)$-approximate current-state opaque over time horizon $n \!\in\!\! \mathbb{N}$ if and only if for any $x_0 \!\in\! X_0$
$$
\mathbb{P}^{\max}(\xi^C_{x_0}\models \lozenge^{\leq n} X_C^{bad}) \leq 1-\lambda.
$$
\end{theorem}
\begin{IEEEproof}
This theorem can be proved by the same reasoning for Theorem~\ref{thm:veri_ini}.
\end{IEEEproof}

\begin{remark} \label{remark:computation}
The precise values of $\mathbb{P}^{\max}(\xi^I_{x_0}\models \lozenge^{\leq n} X_I^{bad})$  and $\mathbb{P}^{\max}(\xi^C_{x_0}\models \lozenge^{\leq n} X_C^{bad})$ can be computed by following the value iteration processes in Lemma \ref{lemma:LP}, using  existing computational tools such as \texttt{MDPtoolbox}~\cite{MDPToolbox}.
Note that since the estimators $\Sigma_{I,obs}$ and $\Sigma_{C,obs}$ are \emph{finite} gMDPs, the values of $\mathbb{P}^{\max}(\xi^I_{x_0}\models \lozenge^{\leq n} X_I^{bad})$ and $\mathbb{P}^{\max}(\xi^C_{x_0}\models \lozenge^{\leq n} X_C^{bad})$ can be computed in \emph{polynomial time} in the size of the estimators. As a consequence, the question of whether 
a finite gMDP is approximate initial-state/current-state opaque for some $\lambda \in [0, 1)$ is \emph{decidable} in polynomial time.
\end{remark}

\vspace{-0.2cm}

\section{Approximate Opacity-preserving Stochastic Simulation Relations}\label{Sec:systemrelations}

In the previous sections, we introduced notions of approximate opacity for gMDPs, and presented opacity verification approaches for finite gMDPs.  
However, the verification of approximate opacity directly on gMDPs with infinite state sets is infeasible. Therefore, in the next sections, we propose an abstraction-based approach for the opacity verification problem for infinite gMDPs by leveraging new notions of system relations between a gMDP and its finite abstraction. 

To this end, in this section, we introduce new notions of approximate opacity-preserving stochastic simulation relations based on lifting for gMDPs, and then show their usefulness in the verification of approximate opacity for gMDPs.  
Let us first recall the notion of $\delta$-lifted relation proposed in \cite{haesaert2017verification}.
\begin{definition}($\delta$-lifted relation \cite{haesaert2017verification}) \label{def:liftedrelation}
	Let $X$ and $\hat X$ be two sets associated with the corresponding measurable spaces $(X, \mathcal B(X))$ and $(\hat X, \mathcal B(\hat X))$. Consider a relation $R \subseteq X \times \hat{{X}}$ with $R \in \mathcal B(X \times \hat{{X}})$. We denote by $\bar R_{\delta} \subseteq \mathcal{P}(X,\mathcal B( X)) \times \mathcal{P}(\hat X,\mathcal B(\hat X))$ the corresponding \emph{$\delta$-lifted relation} if there exists a probability measure $L$, referred to as a \emph{lifting}, for the probability space $(X \times \hat{{X}}, \mathcal B(X \times \hat{{X}}), L)$ such that $(\Phi, \Theta) \in \bar R_{\delta}$, denoted by $\Phi\bar R_{\delta}\Theta$,  if and only if 
	\begin{enumerate}
		\item  $\forall A \in \mathcal B(X)$, $L(A \times \hat X) = \Phi(A)$; 
		\item $\forall \hat A \in \mathcal B(\hat X)$, $L(X \times \hat A) = \Theta(\hat A)$; 
		\item For the probability space $(X \times \hat{{X}}, \mathcal B(X \times \hat{{X}}), L)$, it holds that $xR\hat x$ with probability at least $1-\delta$, equivalently, $L(R) \in [1-\delta,1]$.
 \end{enumerate}
\end{definition}
Given a relation $R \subseteq X \times \hat{{X}}$, the above definition provides conditions under which the relation $R$ can be lifted to a relation $\bar R_{\delta}$ that relates probability measures over state sets $X$ and $\hat{{X}}$. 
The third condition in Definition \ref{def:liftedrelation} requires that the probability measure $L$ assigns a probability of at least $1-\delta$ to the set of state pairs in the relation $R$.

In the next subsections, we introduce the new notions of \emph{approximate opacity-preserving stochastic simulation relations} based on $\delta$-lifted relation as in the above definition. The stochastic simulation relations can be used to capture the similarities between a gMDP and its finite abstraction.

\vspace{-0.2cm}
\subsection{Initial-State Opacity-Preserving Stochastic Simulation Relation}	
First, we introduce a new notion of stochastic simulation relations for preserving approximate initial-state opacity. 

\begin{definition} \label{InitialRinter}
(\textbf{Approximate Initial-State Opacity-Preserving Stochastic Simulation Relation})
	Consider two  gMDPs $\Sigma\!=\!(X,X_0,X_S,U,T,Y, h)$ and $\hat\Sigma \!=\!(\hat{{X}},\hat{{X}}_0,\hat{{X}}_S,\hat{{U}},$ $\hat T,\hat{{Y}},\hat h)\!$ with the same output sets $Y = \hat Y$. 
 System $\hat{\Sigma}$ is \emph{$(\epsilon, \delta)$-initial-state opacity-preservingly ($(\epsilon, \delta)$-InitSOP)} simulated by  $\Sigma $, i.e.,  $\hat\Sigma \preceq_{{\mathcal {I}}^{\epsilon}_\delta} {\Sigma} $, if there exists a relation 
 $R_x \subseteq X \times \hat{{X}}$ and  
 Borel measurable stochastic kernels $L_T(\cdot|x,\hat x, \hat u)$ and $L_{\hat T}(\cdot|x,\hat x, u)$ on $ X \times \hat{{X}}$ such that 
	\begin{enumerate} 
		\item 
  \begin{enumerate} 
		\item   $\forall x_0 \in {X}_0 \cap {X}_S$, $\exists \hat x_0 \in \hat { X}_0 \cap \hat { X}_S$, s.t. $(x_0,\hat x_0) \in R_x$;
  \item  $\forall \hat x_0 \in \hat {X}_0 \setminus \hat {X}_S$, $\exists x_0 \in {X}_0 \setminus {X}_S$, s.t. $(x_0,\hat x_0) \in R_x$;
  \end{enumerate}	
		\item  $\forall (x, \hat{x}) \in R_x$, $\Vert h(x)-\hat h(\hat x)\Vert \leq \epsilon$;
		\item  $\forall (x, \hat{x}) \in R_x$, we have
 \begin{enumerate} 
		\item $\forall \hat u \in \hat{{U}}$, $\exists u \in U$, s.t. $T(\cdot|x,u)$$\bar R_{\delta}$$\hat T(\cdot|\hat x,\hat u)$ with lifting $L_T(\cdot|x,\hat x, \hat u)$; 
		\item $\forall u\in U$, $\exists \hat u \in \hat{{U}}$, s.t. $T(\cdot|x,u)$$\bar R_{\delta}$$\hat T(\cdot|\hat x,\hat u)$ with lifting $L_{\hat T}(\cdot|x,\hat x, u)$.
  \end{enumerate}
	\end{enumerate}
\end{definition} 


The third condition of Definition~\ref{InitialRinter} implies that for any $\hat u \in \hat U$, there exists $u \in U$, and vice versa, such that the state probability measures are in the lifted relation after one state transition for any $(x, \hat x) \in R_x$. In other words, the $\delta$-lifted relation $\bar R_{\delta}$ requires that the state pairs remain in the relation $R_x$ in the next time step with a probability of at least $1-\delta$ if they are in the relation $R_x$ at the current time step.  
This new notion of stochastic simulation relation can be used to relate two gMDPs in terms of preserving approximate initial-state opacity. In an $(\epsilon, \delta)$-InitSOP stochastic simulation relation, the parameter $\delta$ quantifies the distance in probability between gMDPs, and $\epsilon$ is used to capture the closeness of output trajectories of two gMDPs. 
If $\hat\Sigma \preceq_{{\mathcal {I}}^{\epsilon}_\delta} {\Sigma}$ with $\hat{\Sigma}$ being a (finite) gMDP, $\hat{\Sigma}$ is called an InitSOP (finite) abstraction of system $\Sigma$.

Before showing the usefulness of this relation, let us denote for any measurable set $E$ over the output trajectories of a gMDP $\Sigma$ with $E \subseteq \mathcal{H}^n$, the \emph{$\epsilon$-expansion} and \emph{$\epsilon$-contraction} of set $E$, denoted by $E^{\epsilon}$ and $E^{-\epsilon}$, respectively. 
In particular, for any $\epsilon>0$, $E^{\epsilon}$ and $E^{-\epsilon}$ are the $\epsilon$ neighborhoods of set $E$, where 
$E^{\epsilon}\!$ and $E^{-\epsilon}$  are respectively the smallest measurable set in $\mathcal{H}^n$ containing $E$, and the largest measurable set contained in $E$ satisfying:  %
\begin{align}  \notag
\!\!\!\!\!  E^{\epsilon}=  \{(y_0, \dots, y_n)  \in \mathcal{H}^n: &\exists(\bar y_0, \dots, \bar y_n) \in E, \\ \label{infE} 
&\max_{i \in [0;n]}\Vert \bar y_i \!\! - \!\! y_i \Vert \leq \epsilon \}; \\
	E^{-\epsilon} \!=\! \{(y_0, \dots, y_n)\in \mathcal{H}^n: &\forall (\bar y_0, \dots, \bar y_n) \in \mathcal{H}^n \setminus E , \\ \label{defE}
&\min_{i \in [0;n]}\Vert  \bar y_i -y_i \Vert \geq \epsilon \}.
\end{align}
%
The following proposition provides us a probabilistic closeness guarantee between a gMDP and its InitSOP abstraction.
\begin{proposition} \label{probclosenessinit}
Consider two  gMDPs $\Sigma\!=\!(X,X_0,X_S,$ $U,T,Y, h)$ and $\hat\Sigma \!=\!(\hat{{X}},\hat{{X}}_0,\hat{{X}}_S,\hat{{U}},\hat T,\hat{{Y}},\hat h)\!$ with the same output sets $Y = \hat Y$. 
	If $\hat\Sigma \preceq_{{\mathcal {I}}^{\epsilon}_\delta} {\Sigma}$, then for any $(x_0,\hat x_0) \in R_x$ and any input sequence $\hat\nu$ on $\hat \Sigma$, there exists an input sequence $\nu$ for $\Sigma$, and vice versa, such that for all measurable events $E \subseteq \mathcal{H}^n$ and the $\epsilon$-neighborhoods $E^{\epsilon}$ and $E^{-\epsilon}$ as defined in \eqref{infE} and \eqref{defE}:
	\begin{align} \label{pp2dtscs1}
\mathbb{P}(\hat\zeta_{\hat x_0 \hat\nu} \!\in\!  E^{-\epsilon}) \!- \!\gamma_\delta   \!\leq  \mathbb{P}(\zeta_{x_0\nu} \!\in\! E) \!\leq \mathbb{P}(\hat\zeta_{\hat x_0 \hat\nu}\! \in\! E^{\epsilon}) \!+\! \gamma_\delta,
	\end{align}
where $|\zeta_{x_0\nu}| = |\hat\zeta_{\hat x_0 \hat\nu}| = n$ with constant $\gamma_\delta := 1- (1-\delta)^{n}$.
\end{proposition}

The proof of Proposition \ref{probclosenessinit} is provided in the Appendix.
Before showing the main result of this section, let us define the sets of ``secure output sequences"  with respect to the secure state trajectories $\mathcal{L}_{\mathcal{I}_\varepsilon}^n$ and $\mathcal{L}_{\mathcal{C}_\varepsilon}^n$ as defined in \eqref{secureinitialset} and \eqref{securecurrentset}, respectively:
\begin{align} 	\notag
	\mathcal{H}_{\mathcal{I}_\varepsilon}^n :=\{ (y_0,\dots, y_n) \in \mathcal{H}^n: &\exists  (x_0,\dots, x_n) \in \mathcal{L}_{\mathcal{I}_\varepsilon}^n,   \text{s.t. } \\ \label{secureyini}
	&y_i = h(x_i), \forall i \in [0;n]\}; \\  \notag
	\mathcal{H}_{\mathcal{C}_\varepsilon}^n :=\{  (y_0,\dots, y_n) \in \mathcal{H}^n: & \exists (x_0,\dots, x_n) \in \mathcal{L}_{\mathcal{C}_\varepsilon}^n, \text{s.t. }\\ \label{secureycur}
	& y_i = h(x_i), \forall i \in [0;n]\}, 
\end{align}
where $\mathcal{H}_{\mathcal{I}_\varepsilon}^n$ and $\mathcal{H}_{\mathcal{C}_\varepsilon}^n$ denotes the set of secure output sequences w.r.t. approximate initial-state opacity and approximate current-state opacity, respectively, over time horizon $n \in \mathbb{N}$.
The following lemma will be used later to prove our main results.
\begin{lemma}\label{lemma1}
	Consider a gMDP $\Sigma= (X,X_0,X_S,U,T,$ $Y, h )$, the associated sets of secure state trajectories $L_{\mathcal{I}_\varepsilon}$ and $L_{\mathcal{C}_\varepsilon}$ in \eqref{secureinitialset}-\eqref{securecurrentset}, and the sets of secure output sequences in \eqref{secureyini}-\eqref{secureycur}. 
For any initial state  $x_0$ and input sequence $\nu$, we have
	\begin{align}\label{eqp}
		\mathbb{P}(\xi_{x_0\nu} \in L_{\mathcal{I}_\varepsilon}^n ) \geq \lambda  \Longleftrightarrow \mathbb{P}(\zeta_{x_0\nu} \in \mathcal{H}_{\mathcal{I}_\varepsilon}^n) \geq \lambda;\\ \label{eqp2}
		\mathbb{P}(\xi_{x_0\nu} \in L_{\mathcal{C}_\varepsilon}^n ) \geq \lambda  \Longleftrightarrow \mathbb{P}(\zeta_{x_0\nu} \in \mathcal{H}_{\mathcal{C}_\varepsilon}^n) \geq \lambda.
	\end{align}
\end{lemma}

The proof of Lemma \ref{lemma1} is provided in the Appendix. 
The next theorem presents one of the main results of this paper, which shows 
that the newly introduced stochastic simulation relation in Definition \ref{InitialRinter} can be used for the preservation of approximate initial-state opacity between two gMDPs.

\begin{theorem}\label{thm:InitSOP2}
	Consider two  gMDPs $\Sigma\!=\!(X,X_0,X_S,U,T,$ $Y, h)$ and $\hat\Sigma \!=\!(\hat{{X}},\hat{{X}}_0,\hat{{X}}_S,\hat{{U}},\hat T,\hat{{Y}},\hat h)\!$ with the same output sets $Y = \hat Y$.
Let $\epsilon,\delta\in\mathbb R_{\ge 0}$.
	If $\hat\Sigma \preceq_{{\mathcal {I}}^{\epsilon}_\delta} {\Sigma}$ and  $\gamma_\delta = 1- (1-\delta)^{n} \leq \lambda$,
	then the following holds
	\begin{align}\notag
		&\hat\Sigma \text{ is }(\varepsilon, \lambda)\text{-approximate initial-state opaque} \\ \label{impl1ini} \hspace{-3mm}
		\Rightarrow &\Sigma \text{ is } \!(\varepsilon+2\epsilon,\lambda_\delta) \text{-approximate initial-state opaque},
	\end{align}
where the opacity satisfaction holds within time horizon $n \in \mathbb{N}$ and $\lambda_\delta = (1-\gamma_\delta)(\lambda- \gamma_\delta)$.
\end{theorem}

The proof of Theorem \ref{thm:InitSOP2} is provided in the Appendix. 
This result provides the foundation for verifying approximate initial-state opacity of gMDPs using abstraction-based techniques. 
As gMDPs are equipped with uncountable state spaces as well as metric output spaces, the verification/analysis of properties over gMDPs is generally infeasible.  It is thus promising to approximate these models by simpler ones that are prone to be verified, such as finite gMDPs \cite{haesaert2017verification,lavaei2022automated}. 
In this work, by constructing a finite abstraction $\hat \Sigma$ of the concrete system $\Sigma$ via the InitSOP stochastic simulation relation between them, one can first efficiently verify opacity over the simpler system $\hat{\Sigma}$ by applying the verification approach in Section \ref{sec:veri}, and then carry back the results to the concrete system $\Sigma$ by using Theorem~\ref{thm:InitSOP2}.

The following example illustrates the usefulness of the new notion of InitSOP stochastic simulation relation as presented in Theorem~\ref{thm:InitSOP2}. 

\begin{example}
	\begin{figure}
		\centering
		%
  \includegraphics[width=0.38\textwidth]{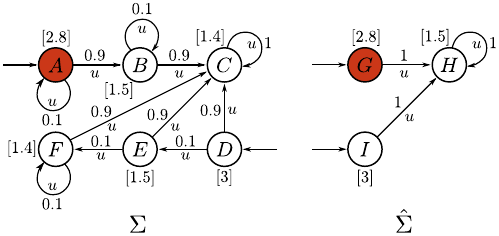}
		\caption{Example of $(\epsilon, \delta)$-InitSOP from $\hat \Sigma$ to $\Sigma$, i.e., $\hat\Sigma \preceq_{{\mathcal {I}}^{\epsilon}_\delta} {\Sigma} $.} 
		\label{exmaple3}
		\vspace{-0.7cm}
	\end{figure}	
Consider two  gMDPs $\Sigma\!=\!(X,X_0,X_S,U,T,$ $Y, h)$ and  $\hat\Sigma \!=\!(\hat{{X}},\hat{{X}}_0,\hat{{X}}_S,\hat{{U}},\hat T,\hat{{Y}},\hat h)\!$ as depicted in Figure \ref{exmaple3}, where $X =\{A,B,C,D,E,F\}$, $X_0 = \{A,D\}$, $X_S = \{A\}$;
$\hat{X} = \{G,H,I\}$, $\hat{X}_0 = \{G, I\}$, $\hat{X}_S = \{G\}$, $U = \hat{{U}} = \{u\}$. 
All secret states are marked in red, and the output of each state is specified by the value associated with it. 
	Let us consider the following relation
	$R_{x}=\{(A,G),(B,H),(C,H),(E,H),(F,H),(D,I)\}$. We claim that $R_{x}$ is an
$(\epsilon,\delta)$-initial-state opacity-preserving  stochastic simulation relation  from $\hat \Sigma$ to $\Sigma$ with $\epsilon=0.1$ and $\delta = 0.1$.  
This is shown by checking the conditions in Definition~\ref{InitialRinter} as follows. 
First, for the secret initial state $A \in X_0 \cap X_S$, there exists  $G \in \hat X_0 \cap \hat{{X}}_S$, s.t. $(A,G) \in R_x$. Similarly, for the non-secret initial state $I \in \hat X_0 \setminus \hat{{X}}_S$, there exists $D \in X_0 \setminus X_S$ s.t. $(D,I) \in R_x$. Thus, condition 1 in Definition \ref{InitialRinter} holds. 
Also, one can check that $\Vert h(x)- \hat h(\hat x) \Vert \leq \epsilon$ holds, with $\epsilon=0.1$, for all $(x, \hat x) \in  R_x$, which implies that condition 2 in Definition \ref{InitialRinter} also holds. 
Moreover, one can check that  conditions 3a) and 3b)  hold for any $(x, \hat x) \in  R_x$ as well. As an example, for $(A,G) \in R_x$ and  $u \in U$, there exists $u \in \hat U$, s.t., with probability $1-\delta = 0.9$, the succeeding states are $B$ and $H$ with $(B,H)$ being in relation $R_x$. 
One can conclude that $R_x$ is an $(\epsilon,\delta)$-initial-state opacity-preserving  stochastic simulation relation  from $\hat \Sigma$ to $\Sigma$ with $\epsilon=0.1$ and $\delta = 0.1$, i.e., $\hat\Sigma \preceq_{{\mathcal {I}}^{0.1}_{0.1}}{\Sigma}$.
By applying the verification approach in Section \ref{subsec:veriinitial}, we can easily check that $\hat \Sigma$ is $(0.2,1)$-approximate initial-state opaque within any finite time horizon $n \in \mathbb{N}$. Therefore, according to Theorem \ref{thm:InitSOP2}, we conclude that is $(0.4,0.81^n)$-approximate initial-state opaque within finite time horizon $n \in \mathbb{N}$, without needing to apply the verification approach to $\Sigma$ directly.
	\hfill$\diamond$ 
\end{example}

\vspace{-0.2cm}

\subsection{Current-State Opacity-Preserving Stochastic Simulation Relation}	
Here, we introduce a new notion of approximate current-state opacity-preserving stochastic simulation relations. 

\begin{definition} \label{CurRinter}
(\textbf{Approximate Current-State Opacity-Preserving Stochastic Simulation Relation})
	Consider two  gMDPs $\Sigma\!=\!(X,X_0,X_S,U,T,Y, h)$ and $\hat\Sigma \!=\!(\hat{{X}},\hat{{X}}_0,\hat{{X}}_S,\hat{{U}},$ $\hat T,\hat{{Y}},\hat h)\!$ with the same output sets $Y = \hat Y$. 
  System $\hat{\Sigma}$ is \emph{$(\epsilon, \delta)$-current-state opacity-preservingly ($(\epsilon, \delta)$-CurSOP)} simulated by  $\Sigma $, i.e.,  $\hat\Sigma \preceq_{{\mathcal {C}}^{\epsilon}_\delta} {\Sigma} $, if there exists a relation 
  $R_x \subseteq X \times \hat{{X}}$ and Borel measurable stochastic kernels $L_T(\cdot|x,\hat x, \hat u)$ and $L_{\hat T}(\cdot|x,\hat x, u)$ on $ X \times \hat{{X}}$ such that 
	\begin{enumerate}
		\item 
  \begin{enumerate}
	\item $\forall x_0 \in {X}_0$, $\exists \hat x_0 \in \hat {X}_0$, s.t. $(x_0,\hat x_0) \in R_x$;
 	\item $\forall \hat x_0 \in \hat {X}_0$, $\exists  x_0 \in {X}_0$, s.t. $(x_0,\hat x_0) \in R_x$;
  \end{enumerate}
		\item $\forall (x, \hat{x}) \in R_x$, $\Vert h(x)-\hat h(\hat x)\Vert \leq \epsilon$;
		\item  $\forall(x, \hat{x}) \in R_x$, we have\footnote{We denote  $x^+ := f(x,u, \varsigma)$ and $\hat x^+ := \hat f(\hat x, \hat u, \varsigma)$ for the sake of brevity.} 
  \begin{enumerate}
      \item $\forall u\in U$, $\exists \hat u \in \hat{{U}}$, s.t. $T(\cdot|x,u)$$\bar R_{\delta}$$\hat T(\cdot|\hat x,\hat u)$ with lifting $L_{\hat T}(\cdot|x,\hat x, u)$; 
	\item $\forall u\in U$, s.t. $x^+ \in X_S$, $\exists \hat u \in \hat{{U}}$, s.t. $\hat x^+ \in \hat X_S$ and $T(\cdot|x,u)$$\bar R_{\delta}$$\hat T(\cdot|\hat x,\hat u)$ with lifting $L_{\hat T}(\cdot|x,\hat x, u)$; 
	\item  $\forall \hat u \in \hat{{U}}$, $\exists u \in U$, s.t. $T(\cdot|x,u)$$\bar R_{\delta}$$\hat T(\cdot|\hat x,\hat u)$ with lifting $L_T(\cdot|x,\hat x, \hat u)$; 
	\item  $\forall \hat u \in \hat{{U}}$,  s.t. $\hat x^+ \in \hat X \setminus \hat X_S$, $\exists u \in U$, s.t. $x^+ \in X \setminus X_S$ and $T(\cdot|x,u)$$\bar R_{\delta}$$\hat T(\cdot|\hat x,\hat u)$ with lifting $L_T(\cdot|x,\hat x, \hat u)$.
  \end{enumerate}		
	\end{enumerate}
\end{definition} 
This new notion of stochastic simulation relation can be used to relate two gMDPs in terms of preserving approximate current-state opacity.
If  $\hat\Sigma \preceq_{{\mathcal {C}}^{\epsilon}_\delta} {\Sigma} $ with $\hat{\Sigma}$ being a (finite) gMDP, $\hat{\Sigma}$ is called a CurSOP (finite) abstraction of $\Sigma$. 

The next proposition provides us with a probabilistic closeness guarantee between a gMDP  and its CurSOP abstraction.
\begin{proposition} \label{probclosenesscur}
	Consider two  gMDPs $\Sigma\!=\!(X,X_0,X_S,$ $U,T,Y, h)$ and $\hat\Sigma \!=\!(\hat{{X}},\hat{{X}}_0,\hat{{X}}_S,\hat{{U}},\hat T,\hat{{Y}},\hat h)\!$ with the same output sets $Y = \hat Y$. 
If $\hat\Sigma \preceq_{\mathcal {C}^{\epsilon}_\delta} {\Sigma}$, then for any $(x_0,\hat x_0) \in R_x$ and any input sequence $\hat\nu$ on $\hat \Sigma$, there exists an  input sequence $\nu$ for $\Sigma$, and vice versa, such that for  all measurable events  $E \subseteq \mathcal{H}^n$ and the $\epsilon$-neighborhoods $ E^{\epsilon}$ and $E^{-\epsilon}$ as defined in \eqref{infE}-\eqref{defE}, the inequality \eqref{pp2dtscs1} holds
	with constant $\gamma_\delta = 1- (1-\delta)^{n}$.
\end{proposition}

The proof of this proposition follows similar reasoning as that of Proposition \ref{probclosenessinit} and is omitted here.
By employing the above proposition, we provide the following theorem which shows that the newly introduced stochastic simulation relation in Definition~\ref{CurRinter} can be used for the preservation of approximate current-state opacity between two gMDPs.
%
\begin{theorem}\label{thm:CurSOP}
	Consider two gMDPs $\Sigma\!=\!(X,X_0,X_S,U,T,$ $Y, h)$ and $\hat\Sigma \!=\!(\hat{{X}},\hat{{X}}_0,\hat{{X}}_S,\hat{{U}},\hat T,\hat{{Y}},\hat h)$ with the same output sets $Y = \hat Y$.
	Let $\epsilon,\delta\in\mathbb R_{\ge 0}$.
	If $\hat\Sigma \preceq_{\mathcal {C}^{\epsilon}_\delta} {\Sigma}$ and  $\gamma_\delta = 1- (1-\delta)^{n} \leq \lambda$,
	then the following holds
	\begin{align}\notag
		&\hat\Sigma \text{ is }(\varepsilon, \lambda)\text{-approximate current-state opaque} \\ \label{impl1cur} \hspace{-3mm}
		\Rightarrow &\Sigma \text{ is } \!(\varepsilon+2\epsilon,\lambda_\delta) \text{-approximate current-state opaque},
\end{align} 
where the opacity satisfaction holds within time horizon $n \in \mathbb{N}$ and  $\lambda_\delta = (1-\gamma_\delta)(\lambda- \gamma_\delta)$.
\end{theorem}

The proof of Theorem \ref{thm:CurSOP} is provided in the Appendix. 

\begin{remark}
\bl{Let us remark that the parameters $\epsilon$ and $\delta$ that appeared in Theorems \ref{thm:InitSOP2} and \ref{thm:CurSOP} are related to the opacity-preserving stochastic simulation relations between two gMDPs as in Definitions \ref{InitialRinter} and \ref{CurRinter}, respectively. They are used to specify two different types of precision: $\epsilon$ characterizes the ``closeness" between the output trajectories of two gMDPs in terms of being approximate opaque, while $\delta$ is used to quantify the probability of the pair of state trajectories of two gMDPs remaining in the simulation relation after each time step.}  
\end{remark}

\vspace{-0.2cm}

\section{Finite Abstractions for GMDPs} \label{Sec:abs}
In the previous section, we introduced notions of approximate opacity-preserving stochastic simulation relations and discussed their usefulness in verifying approximate opacity for gMDPs using their abstractions. 
In this section, we will discuss how to establish the proposed simulation relations for gMDPs and their finite abstractions (finite gMDPs). To be specific, we provide conditions under which one can always construct finite abstractions under the desired $(\epsilon, \delta)$-InitSOP (resp. CurSOP) stochastic simulation relations for a (possibly infinite) gMDP.

In the following, we consider a gMDP $\Sigma=(X,X_0,X_S,U,$ $T,Y, h)$ with $X = X_0$, and assume its output map $h$ satisfies the following general Lipschitz assumption 
\begin{align}\label{lipschitz}
	\Vert h(x)-h(x')\Vert\leq \ell(\Vert x-x'\Vert),
\end{align}
for all $x,x'\in X$, where $\ell\in\KK$. 
In addition, the existence of opacity-preserving stochastic simulation relations between $\Sigma$ and its finite abstraction is established under the assumption that $\Sigma$ is \emph{incrementally input-to-state stable} \cite{zamani2014symbolic} as in the following definition.

\begin{definition}\label{Def111}
	A gMDP $\Sigma=\left(X,X_0,X_S,U,T,Y, h\right)$ is called \emph{incrementally input-to-state stable}  ($\delta$-ISS) if there exists a  function $V: X\times X\to \mathbb{R}_{\geq0} $  such that $\forall x, x'\in X$, $\forall u,u'\in U$,   the following conditions hold:
	\begin{align}	\label{Con555a} 
		\underline{\alpha}(&\Vert x-x'\Vert ) \leq V(x,x')\leq \overline{\alpha} (\Vert x-x'\Vert );\\ \notag
	\mathbb{E} \Big[&	V(f(x,u,\varsigma),f(x',u',\varsigma))\big|x,x',u,u' \Big] - V(x,x')\leq \\ \label{Con555b}
	&	-\bar{\kappa}(V(x,x'))+\bar \rho(\Vert u-u'\Vert),
	\end{align}
	for some $\underline{\alpha}, \overline{\alpha}, \bar{\kappa} \in \mathcal{K}_{\infty}$, and $\bar \rho\in\mathcal{K}_\infty\cup\{0\}$.
\end{definition}

The above definition can be seen as a stochastic counterpart of the $\delta$-ISS Lyapunov functions defined in \cite{tran2018convergence} for discrete-time non-stochastic systems. 

In addition, we assume that there exists a function $\gamma\in\mathcal{K}_{\infty}$  such that $V$ satisfies
\begin{equation}\label{Eq65}
	V(x,x')-V(x,x'')\leq \gamma(\Vert x'-x''\Vert), \forall x,x',x'' \in X.
\end{equation}
This assumption is not restrictive in the sense that condition \eqref{Eq65} can be satisfied for any differentiable function $V$ restricted to a compact subset of $X 
\times X$ by applying the mean value theorem \cite{zamani2014symbolic}.

\vspace{-0.3cm}
\subsection{Finite General Markov Decision Processes}	\label{subsec:MDP}
In this subsection, we approximate a gMDP $\Sigma$ with a finite abstraction (finite gMDP) $\hat \Sigma$. 
The construction follows a similar procedure as in  \cite[Algorithm 1]{Lavaeihscc} with some modifications to incorporate the secret information of the system. 

Consider a concrete gMDP $\Sigma\!=\!\left(X,X_0,X_S,U,T,Y, h\right)$, we assume that for the rest of the paper, sets $X$, $X_S$, and $U$ are of the form of finite unions of boxes and $X_{0} = X$. Consider a tuple $q = (\eta, \theta, \mu)$ of parameters, where $0 < \eta \leq \tup{min} \{span(X_S),span({X} \setminus X_S)\}$ is the state set quantization, $0<  \mu  < span ({U})$ is the external input set quantization, and $\theta \in \mathbb{R}_{\ge 0}$ is a design parameter. 

The finite abstraction of $\Sigma$ is a finite gMDP, represented by the tuple $\hat\Sigma \!=\!(\hat{{X}},\hat{{X}}_0,\hat{{X}}_S,\hat{{U}}, \hat T,\hat{{Y}},\hat h)$, where $\hat{{X}} = \hat{X}_0 = [ X]_{\eta}$, $\hat{X}_S = [ X_S^{\theta}]_{\eta}$, $\hat{{U}} = [ U]_{\mu}$, $\hat{{Y}} = \{h(\hat x)|\hat x \in \hat{{X}}\}$, $\hat{h}(\hat x) = h(\hat x)$, $\forall \hat x \in \hat{{X}}$. Intuitively, the state and input sets of $\hat\Sigma$ consist of finitely many representative points, each of which represents an aggregate of continuous states or inputs of the concrete system. 
We denote by ${\it\Pi}_x:X\rightarrow \hat X$ the abstraction map that assigns to any $x\in X$, the representative point $\hat x\in\hat X$ of $x$. Similarly, ${\it\Pi}_u:U\rightarrow \hat U$ is used to denote the abstraction maps for external inputs. Given the quantization parameters, the abstraction maps ${\it\Pi}_x$ and ${\it\Pi}_u$ satisfy the following inequalities 
\begin{equation}
	\label{eq:Pi_delta}
	\Vert {\it\Pi}_x(x)-x\Vert \leq \eta,  \forall x\,\in X;  
	\Vert {\it\Pi}_u(u)-u\Vert \leq \mu, \forall u\,\in U.
\end{equation}
The dynamics $\hat f: \hat X \times \hat U \times V_\varsigma \rightarrow \hat X $ of the abstraction is defined as
$\hat f(\hat x, \hat u, \varsigma) =  {\it {\Pi}}_x(f(\hat x, \hat u, \varsigma))$.
We use $\Xi: \hat X\rightarrow 2^X$ to denote the map that assigns any representative point $\hat x \in X$ to the corresponding partition set represented by $\hat x$, i.e. $\Xi(\hat x)=\{x \in X : {\it\Pi}_x(x) = \hat x\}$.
Consequently, one can compute the transition probability matrix $\hat T$ for $\hat \Sigma$ as $\hat T(\hat x'|\hat x,\hat u) = T(\Xi(\hat x')|\hat x,\hat u)$,
for all $\hat x,\hat x'\in \hat X$, and all $\hat u\in \hat U$. 

\subsection{Establishing Opacity-Preserving Stochastic Simulation Relations}

In the following theorem, we show that for a $\delta$-ISS gMDP $\Sigma$, if the tuple $q = (\eta,\theta,\mu)$ of quantization parameters satisfies certain conditions, then the finite gMDP $\hat \Sigma$ constructed in Subsection~\ref{subsec:MDP} is an $(\epsilon, \delta)$-InitSOP finite abstraction of $\Sigma$.

\begin{theorem} \label{mainthem}
	Consider a $\delta$-ISS gMDP as in Definition \ref{Def111}, associated with function $V$ satisfying \eqref{Con555a}-\eqref{Eq65} with functions $\underline{\alpha}$, $\overline{\alpha}$, $\bar{\kappa}$ $\bar \rho$. For any desired parameters $\epsilon  \in \mathbb{R}_{>0}$, $\mathcal{\delta} \in (0,1]$, let $\hat{\Sigma}$ be a finite abstraction of $\Sigma$ as constructed in Subsection \ref{subsec:MDP}, with the quantization parameters $q = (\eta,0,\mu)$ satisfying 
	\begin{align} \notag
	 \eta \leq\min\{{\gamma^{-1}}(\underline{\alpha}\circ {\ell^{-1}}(\epsilon)\delta -(\mathcal{I}_d&-\bar{\kappa})\circ \underline{\alpha}  \circ {\ell^{-1}}(\epsilon) - \bar \rho(\mu)),  \\    \label{quanti}
  &{\overline{\alpha}^{-1}}\circ\underline{\alpha}  \circ  {\ell^{-1}}(\epsilon)\},
	\end{align}
	then, $\hat\Sigma$ is $(\epsilon, \delta)$-InitSOP simulated by $\Sigma$  via the relation: 
	\begin{align} \label{localsr}
\!\!\!\!	R_{x}  &= \{(x,\hat x) \in X \times \hat X|V(x,\hat x) \leq \underline{\alpha} \circ {\ell^{-1}}(\epsilon) \}.
	\end{align}   
\end{theorem}
The proof of Theorem \ref{mainthem} is provided in the Appendix. 
The following theorem provides another main result of this section, showing that for a $\delta$-ISS gMDP $\Sigma$, if the tuple $q = (\eta,\theta,\mu)$ of quantization parameters satisfies certain conditions, then the finite gMDP $\hat \Sigma$ constructed in Subsection~\ref{subsec:MDP} is an $(\epsilon, \delta)$-CurSOP finite abstraction of $\Sigma$.

\begin{theorem} \label{mainthem_cur}
	Consider a $\delta$-ISS gMDP as in Definition \ref{Def111}, associated with function $V$ satisfying \eqref{Con555a}-\eqref{Eq65} with functions $\underline{\alpha}$, $\overline{\alpha}$, $\bar{\kappa}$ $\bar \rho$. For any desired parameters $\epsilon  \in \mathbb{R}_{>0}$, $\mathcal{\delta} \in (0,1]$, let $\hat{\Sigma}$ be a finite abstraction of $\Sigma$ as constructed in Subsection \ref{subsec:MDP}, with the quantization parameters  $q = (\eta,\theta,\mu)$ satisfying 
	\begin{align}\notag
	 \eta \leq\min\{{\gamma^{-1}}(\underline{\alpha}\circ {\ell^{-1}}(\epsilon)\delta -(\mathcal{I}_d&-\bar{\kappa})\circ \underline{\alpha}  \circ {\ell^{-1}}(\epsilon) - \bar \rho(\mu)),  \\    \label{quanticur}
 & {\overline{\alpha}^{-1}}\circ\underline{\alpha}  \circ  {\ell^{-1}}(\epsilon)\};\\ \label{quanticur2}
	&	{\ell^{-1}}(\epsilon) \leq \theta,
		\end{align}
	then, $\hat\Sigma$ is $(\epsilon, \delta)$-CurSOP simulated by $\Sigma$  via the relation 
	\begin{align} \label{localsr_cur}
		\!\!\!\!	R_{x}  &= \{(x,\hat x) \in X \times \hat X|V(x,\hat x) \leq \underline{\alpha} \circ {\ell^{-1}}(\epsilon) \}.
	\end{align}   
\end{theorem} 
The proof of Theorem \ref{mainthem_cur} is provided in the Appendix. 

\begin{remark}
\bl{As seen from Theorems \ref{mainthem} and \ref{mainthem_cur}, for any desired parameters $\epsilon  \in \mathbb{R}_{>0}$, $\mathcal{\delta} \in (0,1]$, one can always find suitable quantization parameters $\eta$ and $\mu$ to construct a finite abstraction $\hat\Sigma$ that opacity-preservingly simulates the original gMDP $\Sigma$. Recall from Propositions \ref{probclosenessinit} and \ref{probclosenesscur} that the smaller $\delta$ is, the more precise the abstraction $\hat\Sigma$ is in terms of generating similar output trajectories as the original gMDP $\Sigma$. 
However, note that according to conditions \eqref{quanti} and \eqref{quanticur}, a smaller $\delta$ will lead to smaller quantization parameters $\eta$ and $\mu$ for the construction of abstraction $\hat\Sigma$.
This will further result in building a ``fine" abstraction with a state set that is too large, and thus too expensive to compute or store. 
Therefore, from a practical point of view, it is advisable to choose the value of $\delta$ as small as possible, given the available computing power. 
} 
\end{remark}

\begin{remark}
\bl{Observe that the proposed opacity-preserving stochastic simulation relations for the initial state (or the current state) are one-sided because condition 1 in Definition 5.2 (and conditions 3-b and 3-d in Definition 5.7) lacks symmetry. On the other hand, condition 3 in Definition 5.2 (along with conditions 3-a and 3-c in Definition 5.7) exhibits symmetry (two-sided). Thus, in order to establish suitable opacity-preserving stochastic simulation relations, the $\delta$-ISS assumption remains essential for the original gMDPs as detailed in \cite{zamani2014symbolic}.  
} 
\end{remark}

\vspace{-0.2cm}

\section{Case Study}\label{sec:case}



\begin{figure}
	\begin{center}
		\includegraphics[width=0.23\textwidth]{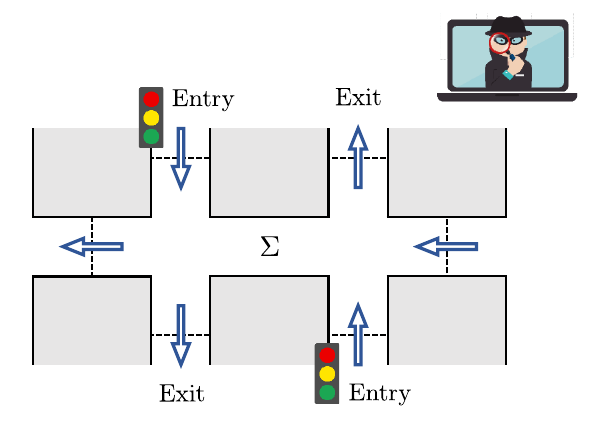}    
		\caption{A road traffic model observed by an outside intruder.} 
		\label{fig:road}
	\end{center}
 \vspace{-0.7cm}
\end{figure}

In this section, we illustrate our results on a road traffic model adapted from \cite{canudas2012graph}, as shown in Fig.~\ref{fig:road}. Here, the traffic flow model is considered as one cell of a road traffic network. 
The cell is assumed to be equipped with at least one measurable entry and one exit. The traffic flow dynamics of the cell is given by
\begin{align*}
\!\!\Sigma:\left\{
\begin{array}{rl}
\mathbf{x}(k+1)= & (1-\frac{\tau v}{l}-e)\mathbf{x}(k) +b\nu(k)+ d\omega(k),\\
\mathbf{y}(k)=&c\mathbf{x}(k),
\end{array}
\right.
\end{align*}
where $\tau$ is the sampling time in hour, $l$ is the length of each cell in kilometers, and
$v$ is the traffic flow speed in kilometers per hour. The state $\mathbf{x}(k)$ of  $\Sigma$ represents the density of the traffic in the vehicle per cell at a specific time instant indexed by $k$. The scalar $b$ denotes the number of vehicles that are allowed to enter the cell during each sampling time controlled by the input signals $\nu(\cdot) \in \{0,1\}$, where $\nu(\cdot) = 1$ (resp. $\nu(\cdot) = 0$) corresponds to green (resp. red) traffic light. The constant $e$ denotes the percentage of vehicles that leave the cell during each sampling time through exits.
The system is considered to be affected by additive noise where $\omega(k)$ is a sequence of independent random variables with standard normal distribution that models environmental uncertainties. The values of the parameters in the system are: $\tau=\frac{30}{60\times 60}$h, $l=1$km, $v=60$km/h, $b=0.5$, $c = 0.1$, $d = 0.1$, and $e = 0.4$.

\begin{figure}
	\begin{center}
		\includegraphics[width=0.30\textwidth]{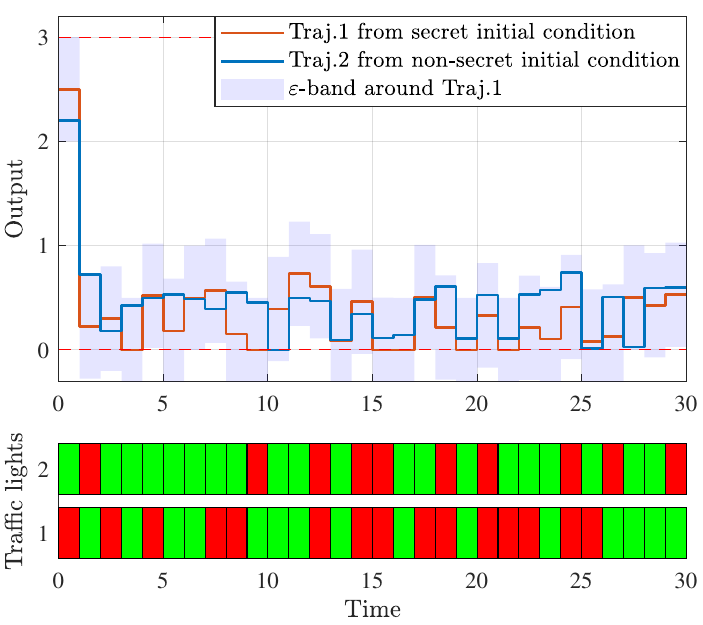}    
		\caption{A state trajectory (Traj.1) initiated from a secret initial condition and its ($\varepsilon$-close) output equivalent trajectory (Traj.2) started from a non-secret initial condition. The shaded area in light blue represents the $\varepsilon$-band around Traj.1.} 
		\label{fig:trajs}
	\end{center}
  \vspace{-0.7cm}
\end{figure}

\begin{figure}
	\begin{center}
		\includegraphics[width=0.30\textwidth]{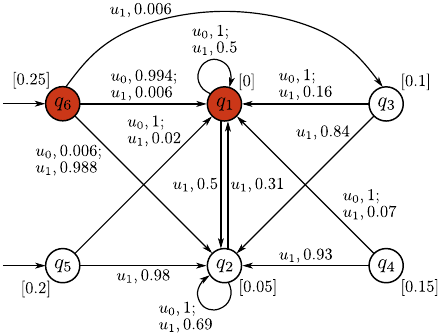}    
		\caption{The constructed finite abstraction $\hat \Sigma$ for the system $\Sigma$ with $\hat\Sigma \preceq_{{\mathcal {I}}^{\epsilon}_\delta} {\Sigma}$. 
  The states marked by red denote secret states; the states marked by sourceless arrows denote initial states; the outputs are specified by the value associated with each state. For each transition, the input and state transition probability are also specified on the arrows.} 
		\label{fig:exampleabs}
	\end{center}
 \vspace{-0.7cm}
\end{figure}

In this example, we consider that the initial density of the traffic model contains critical information that needs to be kept secret (e.g., a high traffic density might indicate activities happening with vehicles gathering in that road cell). 
It is implicitly assumed that there is a malicious intruder (with imprecise measurements captured by $\varepsilon \in \mathbb{R}_{\geq 0}$) who is remotely observing the output behavior of the road cell, intending to carry out an attack. Therefore, the system is interested in verifying its security level in terms of the probability guarantee of maintaining plausible deniability for secret initial conditions. This can be formulated as an ($\varepsilon, \lambda$)-approximate initial-state opacity verification problem as in Definition \ref{opacity} with $\lambda$ characterizing the probability guarantee on the occurrence of secure executions of the system. 
Recall that a secure execution of the system is a state trajectory  $(x_0,\dots, x_n)$, for which there exists a state trajectory generated from a non-secret state such that their output trajectories are indistinguishable for an intruder with measurement precision $\varepsilon$. 
An example of such a pair of trajectories generated by the road traffic model is depicted in Fig.~\ref{fig:trajs}.
Note that the verification of opacity directly on continuous-space control systems is generally infeasible. 

Next, we show the approximate initial-state opacity of $\Sigma$ by applying the abstraction-based verification approach as proposed in this paper. 
We are interested in the state set $X := [0, 3]$, initial set $X_0 := [2, 3]$, and secret set $X_S:= [2.5, 3] \cup [0, 0.5]$. Considering precision parameters $\delta = 0.15$ and $\epsilon = 1$, let us first construct a finite gMDP as the approximate initial-state opacity-preserving abstraction of  $\Sigma$. 
Consider function $V(x, x') = \Vert x - x'\Vert$. One can readily verify that function $V$ satisfies \eqref{Con555a}-\eqref{Eq65} with functions $\underline{\alpha}(s) = \overline{\alpha}(s) = s$, $\bar{\kappa}(s)=(\frac{\tau v}{l}-e)s=0.9s$, $\bar \rho(s) =bs = 0.5s$ and $\ell(s)=0.1s$, which implies that $\Sigma$ is $\delta$-ISS as in Definition~\ref{Def111}. Next, by leveraging Theorem \ref{mainthem}, let us choose $q = (0.5,0,0)$ as the quantization parameters which satisfies \eqref{quanti}, and then construct a finite gMDP following the steps in Subsection~\ref{subsec:MDP}. The obtained finite gMDP  $\hat \Sigma$ is depicted as in Fig.~\ref{fig:exampleabs}. Note that the states $q_1$ to $q_6$ are the representative points in  $\hat \Sigma$ after discretizing the state set using quantization parameter $\eta = 0.5$. Once obtaining this opacity-preserving abstraction $\hat \Sigma$, we apply the verification approach as presented in Section~\ref{sec:veri}, summarized as in the following steps: 1) constructing an $\varepsilon$-approximate initial-state estimator $\Sigma_{I,obs}$ of $\hat \Sigma$; 2) solving the value iteration process for the maximal probabilities of $\Sigma_{I,obs}$ reaching the bad region $X_I^{bad}$ (cf. \eqref{badset}); 3) leveraging Theorem \ref{thm:veri_ini} to conclude the approximate initial-state opacity of $\hat \Sigma$. 
Following these steps, we obtain that $\hat \Sigma$ is $(0.05,1)$-approximate initial-state opaque over any finite time horizon $n \in \mathbb{N}$.  
Note that the verification of initial-state opacity for $\hat \Sigma$ follows exactly the same process as already explained in Example~\ref{example:veri}, and thus, we omit here the detailed process due to lack of space.
Moreover, since $\hat \Sigma$ is $(\epsilon,\delta )$-InitSOP simulated by $\Sigma$ with $\epsilon = 1$ and $\delta = 0.15$, by further leveraging Theorem~\ref{thm:InitSOP2}, we get that $\Sigma$ is $(2.05,0.7225^n)$-approximate initial-state opaque over any time horizon $n \in \mathbb{N}$.

Let us remark that this example serves as an intuitive illustration of the proposed results. We intended to establish a coarse discretization on the state set by choosing a large quantization parameter $\eta$ for the sake of a visualization of the constructed finite gMDP as in Fig.\ref{fig:exampleabs}. 
Instead, one can always resort to existing software tools, e.g., \textsf{Faust}~\cite{FAUST15}, to generate and export a finer abstraction that better approximates the concrete gMDP using smaller quantization parameters. Once the exported abstract model (a finite gMDP) is obtained, one can verify opacity on the finite gMDP by employing the well-known probabilistic model checkers, such as \textsf{PRISM}~\cite{kwiatkowska2002prism}, together with the verification approach presented in Section~\ref{sec:veri}.

 \vspace{-0.2cm}

\section{Conclusion and Discussions}\label{sec:conclusion}

We proposed new notions of approximate initial-state and current-state opacity for continuous-space stochastic control systems. 
This new notion provides a quantifiable measure of opacity, allowing us to quantitatively evaluate the probability of a stochastic system satisfying opacity during each execution. 
We presented novel methods to verify approximate opacity for stochastic systems modeled as finite gMDPs by using value iteration techniques on their approximate initial-state (or current-state) estimators. 
Using newly proposed notions of opacity-preserving simulation relations, an abstraction-based verification approach is then developed to show approximate opacity for infinite gMDPs through their finite abstractions (modeled as finite gMDPs). We also discussed how to construct finite abstractions that preserve approximate opacity for a class of stochastic control systems. 

For future works, we plan to investigate the synthesis of control policies to enforce opacity for gMDPs. \bl{Moreover, we would like to extend the results to other notions of opacity including infinite-step opacity and pre-opacity.
Note that the verification of infinite-step opacity or pre-opacity is much more challenging since we also need to consider how future information can affect our knowledge about the current status of the system. Thus, a new information structure is needed for constructing the estimator which involves the computation of delayed state estimates.  
The challenge in extending our results to verify opacity over infinite time horizons lies in the limitation of the abstraction-based verification technique for \emph{continuous-space} gMDPs. Notice that from Propositions \ref{probclosenessinit} and \ref{probclosenesscur}, the guarantee on the probabilistic distance between a gMDP and its abstraction becomes vacuous when  $n$ goes to infinity. In fact, the problem of building abstractions for stochastic systems for infinite-horizon properties is known to be challenging \cite{lavaei2022automated} due to the conservative nature of Lyapunov-like techniques (simulation relations) and associated error bounds. 
As a potential direction to extend the results to an infinite time horizon, one may consider using similar ideas as proposed in \cite{tkachev2011infinite}, which builds bisimilar finite abstractions for infinite-horizon properties (safety and reachability) by assuming and exploiting some additional structural features, e.g., in terms of the existence of \emph{absorbing sets}\cite{tkachev2011infinite} of the original stochastic system. 
}

\vspace{-0.5cm}
\appendix


{\bf Proof of Lemma~\ref{lemma:1_ini}.} 
Consider an initial state $x_0$ in $\Sigma$ and an input sequence $\nu= (\nu(0), \dots, \nu(n-1))$. For any state trajectory $\xi_{x_0\nu} =(x_0,\dots,x_n)$ in $\Sigma$, by Definition~\ref{def:iniesti} of the approximate initial-state estimator $\Sigma_{I,obs}$, there exists a unique initial-state estimation sequence $((x_0,q_0)\dots(x_n,q_n))$ in $\Sigma_{I,obs}$ generated from $(x_0,q_0)$ under $\nu$. 
Recall that under $\nu= (\nu(0), \dots, \nu(k-1))$, 
the transition probability function $T_I$ in $\Sigma_{I,obs}$ boils down to $\mathbf{P}_I^{\nu(i)}: X_I\times X_I \to [0,1]$ at each time step $i \in [0;n-1]$.  
Thus, $\mathbf{1}^I_{bad} \mathbf{P}_I^{\nu(n-1)}\cdots  \mathbf{P}_I^{\nu(1)} \mathbf{P}_I^{\nu(0)} \mathbf{1}^I_{x_0}$ gives us the probability of a trajectory of $\Sigma_{I,obs}$ starting from initial state $(x_0,q_0)$ under input sequence $\nu$ entering into states in $X_I^{bad}$. 
Now consider that $((x_0,q_0)\dots(x_n,q_n))$ reaches $X_I^{bad}$ at any time step $k \in [0;n]$, then we have  for all $i \in [k;n]$, $(x_i,q_i)\in X_I^{bad}$, and $\{\bar{x}_1\in X: (\bar{x}_1,\bar{x}_2)\in q_i\}\subseteq X_S$ holds by \eqref{badset}.
This means that from time step $k$, the initial-state estimate is a subset of $X_S$, which further implies that when observing the output sequence $(h(x_0),\dots,h(x_n))$ of $\Sigma$, we get $\{x_0'\in X_0: \exists (x_0',\dots, x_n')\in \mathcal{L} \text{ s.t. }\max_{i \in [0;n]} \Vert h(x_i)- h(x_i')\Vert \leq  \varepsilon \}\subseteq X_S$ by \eqref{estimate_ini}. 
Thus, $\xi_{x_0\nu}$ is not initial-state secure, i.e., $\xi_{x_0\nu} \notin \mathcal{L}_{\mathcal{I}_\varepsilon}^n$, from the definition of $\mathcal{L}_{\mathcal{I}_\varepsilon}^n$  in \eqref{secureinitialset}. 
Recall that $T_I(x_I' | x_I, u) = T(x'| x, u)$ also holds by Definition~\ref{def:iniesti}.
Therefore, we get  
$\mathbf{1}^I_{bad} \mathbf{P}_I^{\nu(n-1)}\cdots  \mathbf{P}_I^{\nu(1)} \mathbf{P}_I^{\nu(0)} \mathbf{1}^I_{x_0} 
=   \mathbb{P}(\xi_{x_0\nu} \notin \mathcal{L}_{\mathcal{I}_\varepsilon}^n)=1-\mathbb{P}(\xi_{x_0\nu} \in \mathcal{L}_{\mathcal{I}_\varepsilon}^n)$.  
$\hfill\blacksquare$

{\bf Proof of Lemma~\ref{lemma:opa_ini}.}
Consider an initial state $x_0$ in $\Sigma$ and an input sequence $\nu= (\nu(0), \dots, \nu(n-1))$.  Let $\xi_{x_0\nu} =(x_0,\dots,x_n)$ be an arbitrary state trajectory in $\Sigma$ and $((x_0,q_0)\dots(x_n,q_n))$ be its corresponding initial-state estimation sequence in $\Sigma_{I,obs}$. If \eqref{verifycondition_ini} holds, 
then by leveraging Lemma~\ref{lemma:1_ini}, we get  
$1-\mathbb{P}(\xi_{x_0\nu} \in L_{\mathcal{I}_\varepsilon}^n) \leq  1-\lambda$,
which implies $\mathbb{P}(\xi_{x_0\nu} \in L_{\mathcal{I}_\varepsilon}^n) \geq \lambda$. 
Thus, we conclude that $\Sigma$ is $(\varepsilon, \lambda)$-approximate initial-state opaque  over time horizon $n \in \mathbb{N}$ by Definition~\ref{opacity}. 
$\hfill\blacksquare$

{\bf Proof of Proposition~\ref{probclosenessinit}.}
First note that condition 3 in Definition \ref{InitialRinter} intuitively implies that, for any pair of states $(x,\hat x)$ in relation $R_x$, there always exist inputs such that the succeeding states $(x',\hat x')$ are also in relation $R_x$ with probability at least $1-\delta$. 
	Consequently, by combining the initial conditions in condition 1 of Definition \ref{InitialRinter}, for any pair of initial states that are in relation $(x_0,\hat x_0) \in R_x$ and any control strategy $\nu$, there exists $\hat \nu$ (and vice versa), such that the corresponding state trajectories 
 $\xi_{x_0\nu}$ and ${\hat \xi}_{\hat x_0 \hat \nu}$ satisfy
	\begin{align} \label{prob1}
		\mathbb{P}\{(\xi_{x_0\nu}(k), {\hat \xi}_{\hat x_0 \hat \nu}(k))\in R_x, \forall k \in [0;n]\} \geq (1-\delta)^{n}.
	\end{align}
	Now, let us consider the following events for any set $E \subseteq \mathcal{H}^n$:
	 $\mathcal E_1:=\{\hat\zeta_{\hat x_0 \hat \nu} \in E^{-\epsilon}\}, \mathcal E_2:=\{\zeta_{x_0\nu} \in E\}, 	\mathcal E_3:=\{(\xi_{x_0\nu}(k), {\hat \xi}_{\hat x_0 \hat \nu}(k))\in R_x, \forall k \in [0;n]\}$,	 
 where  $\zeta_{x_0\nu}$ and ${\hat \zeta}_{\hat x_0 \hat \nu}$ are the corresponding output trajectories, and  $E^{\epsilon}$ and $E^{-\epsilon}$ are  defined in \eqref{infE} and \eqref{defE}, respectively.
	From condition 2 of Definition \ref{InitialRinter}, we get $\mathcal E_1 \cap \mathcal E_3 \Longrightarrow \mathcal E_2$, and thus $\mathbb{P}(\mathcal E_1 \cap \mathcal E_3) \leq \mathbb{P}(\mathcal E_2)$. This further implies that the following inequality holds
 	$\mathbb{P}(\bar {\mathcal E}_2)  = 1 - \mathbb{P}(\mathcal E_2)  \leq 1- \mathbb{P}(\mathcal E_1 \cap \mathcal E_3) = \mathbb{P}(\bar {\mathcal E}_1 \cup \bar {\mathcal E}_3)   \leq \mathbb{P}(\bar {\mathcal E}_1) + \mathbb{P}(\bar {\mathcal E}_3)$, 
	where $\bar {\mathcal E}_1$, $\bar {\mathcal E}_2$ and $\bar {\mathcal E}_3$ denote the complements of events ${\mathcal E}_1$, ${\mathcal E}_2$ and ${\mathcal E}_3$, respectively. Since we have by \eqref{prob1}, $\mathbb{P}(\bar {\mathcal E}_3) \leq 1-(1-\delta)^{n}$, then we get 
	$\mathbb{P}(\bar {\mathcal E}_2)\leq \mathbb{P}(\bar {\mathcal E}_1) + 1-(1-\delta)^{n}  
		\Longrightarrow  1 - \mathbb{P}({\mathcal E}_2) \leq 1 -\mathbb{P}({\mathcal E}_1)\!+\! 1\!-\!(1-\delta)^{n}  
		\Longrightarrow  \mathbb{P}({\mathcal E}_1) \leq \mathbb{P}({\mathcal E}_2) + 1-(1-\delta)^{n},
	$
	which gives us $\mathbb{P}(\hat\zeta_{\hat x_0 \hat \nu} \in E^{-\epsilon})\! -\! \gamma_\delta \!  \leq \! \mathbb{P}(\zeta_{x_0\nu} \in E) $. The proof of $\mathbb{P}(\zeta_{x_0\nu} \in E) \leq \mathbb{P}(\hat\zeta_{\hat x_0 \hat \nu} \in E^{\epsilon}) \!+\! \gamma_\delta$ uses similar reasoning and is omitted here.
$\hfill\blacksquare$

{\bf Proof of Lemma~\ref{lemma1}.}
	We provide the proof for \eqref{eqp}. The proof of  \eqref{eqp2}  can be derived similarly. 
 Let us consider events $\mathcal E_1 := \{\xi_{x_0\nu} \in L_{\mathcal{I}_\varepsilon}^n\}$ and $\mathcal E_2 := \{\zeta_{x_0\nu} \in \mathcal{H}_{\mathcal{I}_\varepsilon}^n\}$.	Note that the proof of \eqref{eqp} can be done by showing ${\mathcal E_1} \Longleftrightarrow \mathcal E_2$. 
	Consider an arbitrary initial state $x_0$, an arbitrary input sequence $\nu$, a finite state run $\xi_{x_0\nu} =(x_0,\dots, x_n)$ and the corresponding output trajectory $\zeta_{x_0\nu} = (y_0,\dots, y_n)$ with $y_i = h(x_i)$. 
	
	First, let us prove that ${\mathcal E}_1 \Longrightarrow \mathcal E_2$. Suppose
	$\xi_{x_0\nu} =(x_0,\dots, x_n) \in L_{\mathcal{I}_\varepsilon}^n$. Then, 
	by \eqref{secureyini}, we get $\zeta_{x_0\nu} \in \mathcal{H}_{\mathcal{I}_\varepsilon}^n$ holds readily, which gives us ${\mathcal E}_1 \Longrightarrow \mathcal E_2$. 	
	Next, we show $\mathcal E_2  \Longrightarrow {\mathcal E}_1$.  
	Suppose $\zeta_{x_0\nu} =(y_0,\dots, y_n) \in \mathcal{H}_{\mathcal{I}_\varepsilon}^n$,
	we have by \eqref{secureyini}
	\begin{align}\label{pp1}
		\exists \xi_{x_0'\nu'}=(x_0',\dots, x_n') \in L_{\mathcal{I}_\varepsilon}^n, y_i = h(x_i'), \forall i \in [0;n].
	\end{align}
	Additionally by \eqref{secureinitialset}, there exists $(x_0'',\dots, x_n'') \in \mathcal{L}^n$ s.t.
	\begin{align} \label{pp2}
	  x_0'' \in X_0\!\setminus\! X_S, \text{ and } 		\max_{i \in [0;n]}\Vert h(x_i')-\!h(x_i'') \Vert \leq  \varepsilon.
	\end{align}
	Combining \eqref{pp2} with \eqref{pp1}, we have $\Vert y_i-\!h(x_i'') \Vert \leq \varepsilon, \forall i \in [0;n]$. Therefore, we have for $\xi_{x_0\nu}=(x_0,\dots, x_n)$, $\exists x_0'' \in X_0\!\setminus\! X_S$, s.t. $ 
	\max_{i \in [0;n]}\Vert y_i-\!h(x_i'')\Vert\! \leq \!\varepsilon$.
Thus, it follows that $\xi_{x_0\nu} \in  L_{\mathcal{I}_\varepsilon}^n$ holds as well, which further implies $\mathcal E_2  \Longrightarrow {\mathcal E}_1$. 
$\hfill\blacksquare$

{\bf Proof of Theorem~\ref{thm:InitSOP2}.}
First note that for a non-secret initial-state $x_0 \in {X}_0 \setminus {X}_S$, any trajectory from $x_0$ satisfies $\xi_{x_0\nu} \in \mathcal{L}_{\mathcal{I}_{\varepsilon}}^n$, which directly implies  
that  $\mathbb{P}(\xi_{x_0\nu} \in \mathcal{L}_{\mathcal{I}_{\varepsilon}}^n) = 1$ for any $\varepsilon \in \mathbb R_{\ge 0}$. 
Now, consider an arbitrary initial state $x_0 \in {X}_0 \cap {X}_S$ and a random state trajectory $\xi_{x_0\nu} = (x_0, \dots, x_n)$  with length $n$ 
 generated from $x_0$ under an input sequence $\nu$ in system $\Sigma$.   
	Since $\hat\Sigma \preceq_{{\mathcal {I}}^{\epsilon}_\delta} {\Sigma}$, by
	condition 1a) in Definition \ref{InitialRinter}, for $x_0 \in {X}_0 \cap {X}_S$ in $\Sigma$, there exists $\hat x_0 \in \hat {X}_0 \cap \hat { X}_S$ in $\hat \Sigma$, such that $(x_0,\hat x_0) \in R_x$. Consider this pair of initial states $(x_0,\hat x_0) \in R_x$, by Proposition \ref{probclosenessinit} together with condition 3a) in Definition \ref{InitialRinter}, there exists an input sequence  $\hat\nu$ in $\hat \Sigma$, s.t. 
	given any $E \subseteq \mathcal{H}^n$ and the $\epsilon$-neighborhoods $ E^{\epsilon}$ and $E^{-\epsilon}$ as in \eqref{infE} and \eqref{defE}, we have
	\begin{equation}\label{pp2dtscs2}
 \mathbb{P}(\hat\zeta_{\hat x_0 \hat\nu} \!\in\!  E^{-\epsilon}) \!- \!\gamma_\delta   \!\leq  \mathbb{P}(\zeta_{x_0\nu} \!\in\! E) \!\leq \mathbb{P}(\hat\zeta_{\hat x_0 \hat\nu}\! \in\! E^{\epsilon}) \!+\! \gamma_\delta,
	\end{equation}
	with constant $\gamma_\delta = 1- (1-\delta)^{n}$, where $\zeta_{x_0\nu}$ and $\hat\zeta_{\hat x_0 \hat\nu}$ denotes the output sequence corresponding to the state trajectory $\xi_{x_0\nu}$ in $\Sigma$ and the state trajectory $\hat\xi_{\hat x_0 \hat\nu}$  in $\hat \Sigma$, respectively. Now notice that since $\hat\Sigma$ is $(\varepsilon, \lambda)$-approximate initial-state opaque within time horizon $n \in \mathbb{N}$,  we have from Definition~\ref{opacity} that $\mathbb{P}(\hat\xi_{\hat x_0\hat\nu} \in \hat{\mathcal{L}}_{\mathcal{I}_\varepsilon}^n) \geq \lambda$, where $\hat{\mathcal{L}}_{\mathcal{I}_\varepsilon}^n$ is the set of initial-state secure state trajectories in $\hat \Sigma$ as defined in \eqref{secureinitialset}.  Thus, by Lemma~\ref{lemma1}, the corresponding output run $\hat \zeta_{\hat x_0 \hat \nu} = (\hat y_0,\dots, \hat y_n)$ satisfies
	\begin{align}\label{eq1}
		\mathbb{P}(\hat \zeta_{\hat x_0 \hat \nu} \in \hat{\mathcal{H}}_{\mathcal{I}_\varepsilon}^n) \geq \lambda,
	\end{align}
	where $\hat{\mathcal{H}}_{\mathcal{I}_\varepsilon}^n$ is the set of secure output sequences in $\hat \Sigma$ as  defined in \eqref{secureyini}, i.e., 
	\begin{align} \notag
		\hat{\mathcal{H}}_{\mathcal{I}_\varepsilon}^n =  \{(\hat y_0,\dots, \hat y_n) \in \hat{\mathcal{H}}^n: &\exists (\hat x_0', \dots,\hat x_n') \in \hat{\mathcal{L}}_{\mathcal{I}_\varepsilon}^n, \text{s.t. } \\ \label{H1} 
		 &  \hat y_i =  \hat h(\hat x_i'), \forall i \in [0;n]\}.
	\end{align} 
Combining  \eqref{pp2dtscs2} and \eqref{eq1}, we have 
\begin{align} \label{Prob_ep}
	\mathbb{P}(\zeta_{x_0\nu} \in  \hat{\mathcal{H}} ^{\epsilon}_{\mathcal{I}_\varepsilon} ) \geq \mathbb{P}(\hat\zeta_{\hat x_0 \hat\nu} \in \hat{\mathcal{H}}_{\mathcal{I}_\varepsilon}^n) - \gamma_\delta   \geq  \lambda- \gamma_\delta,
\end{align}
where $\hat{\mathcal{H}} ^{\epsilon}_{\mathcal{I}_\varepsilon}$ is $\epsilon$-expansion of  $\hat{\mathcal{H}}_{\mathcal{I}_\varepsilon}^n$ as defined in \eqref{infE}, which can be further rewritten as
\begin{align}\notag 
\hat{\mathcal{H}} ^{\epsilon}_{\mathcal{I}_\varepsilon}  \stackrel{\eqref{infE}}{=}  \{(\hat{\bar{y}}_0,\dots, \hat{\bar{y}}_n) \in \hat{\mathcal{H}}^n:\exists
(\hat y_0,\dots, \hat y_n) \in 	\hat{\mathcal{H}}_{\mathcal{I}_\varepsilon}^n&, \text{s.t. }  \\ \notag 
 \max_{i \in [0;n]} \Vert \hat y_i - \hat{\bar{y}}_i \Vert \leq \epsilon &  \} \\ \notag 
\stackrel{\eqref{H1}}{=} \{(\hat{\bar{y}}_0,\dots, \hat{\bar{y}}_n) \in \hat{\mathcal{H}}^n: 
\exists (\hat x_0', \dots,\hat x_n') \in \hat{\mathcal{L}}_{\mathcal{I}_\varepsilon}^n& , \text{s.t. }\\ 
\label{abssecureoutput2} 
\max_{i \in [0;n]} \Vert \hat{\bar{y}}_i - \hat h(\hat x_i') \Vert \leq \epsilon &\}.
\end{align}

Next, in order to show the approximate initial-state opacity of system $\Sigma$, let us consider the set of secure output runs within time horizon $n \in \mathbb{N}$ in $\Sigma$ w.r.t.  the constant $(\varepsilon+2\epsilon) \geq 0$:
	\begin{align} \notag
 \mathcal{H}_{\mathcal{I}_{(\varepsilon+2\epsilon)}}^n = \{(y_0,\dots, y_n) \in \mathcal{H}^n : \exists  (x_0,\dots, x_n)& \in \mathcal{L}_{\mathcal{I}_{(\varepsilon+2\epsilon)}}^n,  \\  \label{2e}
 \text{s.t. }  y_i = h(x_i), \forall i \in &[0;n]\}.
\end{align}	

Consider the trajectory $(\hat x_0', \dots,\hat x_n') \in \hat{\mathcal{L}}_{\mathcal{I}_\varepsilon}^n$ in \eqref{abssecureoutput2}.  
By \eqref{secureinitialset}, we get that 
there exists 
$(\hat x_0'', \dots,\hat x_n'') \in \hat{\mathcal{L}}^n$  such that 
\begin{align} \label{ieq1} 
	 \hat  x_0'' \in \hat  X_0\!\setminus\! \hat  X_S, \text{ and } 		\max_{i \in [0;n]}\Vert \hat h(\hat x_i')-\!\hat h(\hat x_i'') \Vert \leq  \varepsilon.
\end{align}

Note that by conditions 1b) and 3b) in Definition~\ref{InitialRinter}, for $\hat x_0'' \in \hat  X_0\!\setminus\! \hat  X_S$ in $\hat \Sigma$, there exists $x_0' \in X_0 \setminus\! X_S$ in $\Sigma$ with $(x_0',\hat x_0'') \in R_x$, such that the trajectory $(x_0', \dots, x_n')$ from $x_0'$ satisfies $(x_i', \hat x_i'')\in R_x$, $\forall i \in [0;n]$ with probability at least $(1-\delta)^{n}$, i.e., 
$ \mathbb{P}\{(x_i', \hat x_i'')\in R_x, \forall i \in [0;n]\} \geq (1-\delta)^{n}$. 
Let us consider again the 
random trajectory $\xi_{x_0\nu}$ in $\Sigma$ and its corresponding output trajectory $\zeta_{x_0\nu}$, and define the events: 
$\mathcal E_1:=\{(x_i', \hat x_i'')\in R_x, \forall i \in [0;n]\}$; $\mathcal E_2:=\{\zeta_{x_0\nu} \in  \hat{\mathcal{H}} ^{\epsilon}_{\mathcal{I}_\varepsilon} \}$; and $\mathcal E_3:=\{\zeta_{x_0\nu} \in \mathcal{H}_{\mathcal{I}_{(\varepsilon+2\epsilon)}}^n  \}$. 
Note that $\mathcal E_1$ further implies that 
\begin{align}\label{ieq2} 
    \max_{i \in [0;n]} \Vert h(x_i') - \hat h(\hat x_i'') \Vert \leq \epsilon,
\end{align} 
by condition 2 in Definition~\ref{InitialRinter}. 
Now consider $\zeta_{x_0\nu} =(y_0,\dots, y_n)$ with $\mathcal E_2:=\{\zeta_{x_0\nu} \in  \hat{\mathcal{H}} ^{\epsilon}_{\mathcal{I}_\varepsilon} \}$ holds,
by leveraging the above implications from $\mathcal E_2$, we get that
there exists $(x_0', \dots, x_n')$ with $x_0' \in X_0 \setminus\! X_S$ in $\Sigma$, such that 
$\max_{i \in [0;n]} \Vert h(x_i') - y_i \Vert \leq \varepsilon+2\epsilon$,
where the inequality is obtained by combining  \eqref{abssecureoutput2}, \eqref{ieq1}, and \eqref{ieq2}.
This implies that $\zeta_{x_0\nu} \in \mathcal{H}_{\mathcal{I}_{(\varepsilon+2\epsilon)}}^n$ holds.
Therefore, we obtain that 
$\mathcal E_1 \cap \mathcal E_2 \Longrightarrow \mathcal E_3$ holds, and thus $\mathbb{P}(\mathcal E_1 \cap \mathcal E_2) \leq \mathbb{P}(\mathcal E_3)$. Therefore, we get $\mathbb{P}(\zeta_{x_0\nu} \in 	\mathcal{H}_{\mathcal{I}_{(\varepsilon+2\epsilon)}}^n) \stackrel{\eqref{Prob_ep}}{\geq} (1-\delta)^{n}(\lambda- \gamma_\delta) = (1-\gamma_\delta)(\lambda- \gamma_\delta)$.
By leveraging again \eqref{eqp}, we have $\mathbb{P}(\xi_{x_0\nu} \in L_{\mathcal{I}_{(\varepsilon+2\epsilon)}}^n) \geq (1-\gamma_\delta)(\lambda- \gamma_\delta)$. Thus, we can conclude that $\Sigma$ is  $(\varepsilon+2\epsilon,\lambda_\delta)$-approximate initial-state opaque within time horizon $n \in \mathbb{N}$, with $\lambda_\delta = (1-\gamma_\delta)(\lambda- \gamma_\delta)$.
$\hfill\blacksquare$

{\bf Proof of Theorem~\ref{thm:CurSOP}.}
 Consider an arbitrary initial state $x_0$ and a random state trajectory $\xi_{x_0\nu} = (x_0, \dots, x_n)$ with length $n$  generated from $x_0$ under a control sequence $\nu$ in system $\Sigma$.
	Since $\hat\Sigma \preceq_{\mathcal {C}^{\epsilon}_\delta} {\Sigma}$, by	condition 1a) in Definition \ref{CurRinter}, for $x_0 \in {X}_0$ in $\Sigma$, there exists $\hat x_0 \in \hat {X}_0$ in $\hat \Sigma$, such that $(x_0,\hat x_0) \in R_x$.  
 Now consider this pair of initial states $(x_0,\hat x_0) \in R_x$, by Proposition \ref{probclosenesscur} together with condition 3a) in Definition \ref{CurRinter}, there exists an input sequence  $\hat\nu$ in $\hat \Sigma$, such that 
	given any set $E \subseteq \mathcal{H}^n$ and the $\epsilon$-neighborhoods $ E^{\epsilon}$ and $E^{-\epsilon}$ as defined in \eqref{infE} and \eqref{defE}, we have
	\begin{equation}\label{pp2dtscs2cur}
 \mathbb{P}(\hat\zeta_{\hat x_0 \hat\nu} \!\in\!  E^{-\epsilon}) \!- \!\gamma_\delta   \!\leq  \mathbb{P}(\zeta_{x_0\nu} \!\in\! E) \!\leq \mathbb{P}(\hat\zeta_{\hat x_0 \hat\nu}\! \in\! E^{\epsilon}) \!+\! \gamma_\delta,
	\end{equation}
	with constant $\gamma_\delta = 1- (1-\delta)^{n}$, where $\zeta_{x_0\nu}$ and $\hat\zeta_{\hat x_0 \hat\nu}$ denotes the output sequence corresponding to the state trajectory $\xi_{x_0\nu}$ in $\Sigma$ and the state trajectory $\hat\xi_{\hat x_0 \hat\nu}$  in $\hat \Sigma$, respectively. Now notice that since $\hat\Sigma$ is $(\varepsilon, \lambda)$-approximate current-state opaque within time horizon $n \in \mathbb{N}$,  we have from Definition~\ref{opacity} that $\mathbb{P}(\hat\xi_{\hat x_0\hat\nu} \in \hat L_{\mathcal{C}_\varepsilon}^n) \geq \lambda$, where $\hat{\mathcal{L}}_{\mathcal{C}_\varepsilon}$ is the set of current-state secure state trajectories in $\hat \Sigma$ as defined in~\eqref{securecurrentset}.  Thus, by Lemma~\ref{lemma1}, the corresponding output run $\hat \zeta_{\hat x_0 \hat \nu} = (\hat y_0,\dots, \hat y_n)$ satisfies
	\begin{align}\label{eq1cur}
		\mathbb{P}(\hat \zeta_{\hat x_0 \hat \nu} \in \hat{\mathcal{H}}_{\mathcal{C}_\varepsilon}^n) \geq \lambda,
	\end{align}
	where $\hat{\mathcal{H}}_{\mathcal{C}_\varepsilon}^n$ is the set of secure output sequences in $\hat \Sigma$ as  defined in \eqref{secureycur}, i.e., 
	\begin{align} \notag
		\hat{\mathcal{H}}_{\mathcal{C}_\varepsilon}^n =  \{(\hat y_0,\dots, \hat y_n) \in \hat{\mathcal{H}}^n: &\exists (\hat x_0', \dots,\hat x_n') \in \hat{\mathcal{L}}_{\mathcal{C}_\varepsilon}^n, \text{s.t. } \\ \label{H1cur} 
		 &  \hat y_i =  \hat h(\hat x_i'), \forall i \in [0;n]\}.
	\end{align} 
Combining  \eqref{pp2dtscs2cur} and \eqref{eq1cur}, we have 
$ \mathbb{P}(\zeta_{x_0\nu} \in  \hat{\mathcal{H}} ^{\epsilon}_{\mathcal{C}_\varepsilon} ) \geq \mathbb{P}(\hat\zeta_{\hat x_0 \hat\nu} \in \hat{\mathcal{H}}_{\mathcal{C}_\varepsilon}^n) - \gamma_\delta   \geq  \lambda- \gamma_\delta$,
where $\hat{\mathcal{H}} ^{\epsilon}_{\mathcal{C}_\varepsilon}$ is $\epsilon$-expansion of  $\hat{\mathcal{H}}_{\mathcal{C}_\varepsilon}^n$ as defined in \eqref{infE}, which can be further rewritten as 
\begin{align}\notag 
\hat{\mathcal{H}} ^{\epsilon}_{\mathcal{C}_\varepsilon} \stackrel{\eqref{infE}}{=}  \{(\hat{\bar{y}}_0,\dots, \hat{\bar{y}}_n) \in \hat{\mathcal{H}}^n:\exists
(\hat y_0,\dots, \hat y_n) \in 	\hat{\mathcal{H}}_{\mathcal{C}_\varepsilon}^n&, \text{s.t. }  \\ \notag 
 \max_{i \in [0;n]} \Vert \hat y_i - \hat{\bar{y}}_i \Vert \leq \epsilon &  \} \\ \notag 
\stackrel{\eqref{H1cur}}{=} \{(\hat{\bar{y}}_0,\dots, \hat{\bar{y}}_n) \in \hat{\mathcal{H}}^n: \exists (\hat x_0', \dots,\hat x_n') \in \hat{\mathcal{L}}_{\mathcal{C}_{\varepsilon}}^n& , \text{s.t. }\\ \label{abssecureoutput2cur}
\max_{i \in [0;n]} \Vert \hat{\bar{y}}_i - \hat h(\hat x_i') \Vert \leq \epsilon &\}.
\end{align}

Next, in order to show the approximate current-state opacity of system $\Sigma$, let us consider the set of secure output runs in $\Sigma$ within time horizon $n \in \mathbb{N}$ w.r.t. the constant $\varepsilon+2\epsilon \geq 0$:
	\begin{align} \notag
 \mathcal{H}_{\mathcal{C}_{(\varepsilon+2\epsilon)}}^n = \{(y_0,\dots, y_n) &\in \mathcal{H}^n : \exists  (x_0,\dots, x_n) \in \mathcal{L}_{\mathcal{C}_{(\varepsilon+2\epsilon)}}^n,  \\  \label{2ecur}
 & \text{s.t. } y_i = h(x_i), \forall i \in [0;n]\}.
\end{align}	
Consider the trajectory $(\hat x_0', \dots,\hat x_n') \in \hat{\mathcal{L}}_{\mathcal{C}_{\varepsilon}}^n$ in \eqref{abssecureoutput2cur}. By \eqref{securecurrentset}, we get that 
for any of its prefix $(\hat x_0', \dots,\hat x_k') $, $\forall k \in [0;n]$,
there exists  
$(\hat x_0'', \dots,\hat x_k'') \in \hat{\mathcal{L}}^n$  such that 
\begin{align} \label{ieq1cur} 
	 \hat  x_k'' \in \hat  X\!\setminus\! \hat  X_S, \text{ and } 		\max_{i \in [0;k]}\Vert \hat h(\hat x_i')-\!\hat h(\hat x_i'') \Vert \leq  \varepsilon.
\end{align}
For any of the prefix $(\hat x_0', \dots,\hat x_k') $, by conditions 1b) and 3c)-3d) in Definition~\ref{CurRinter}, there exists 
$x_0' \in X_0$ with $(x_0', \hat x_0') \in R_x$, s.t. the trajectory  $(x_0', \dots, x_k')$ satisfies $(x_i', \hat x_i'')\in R_x$ $\forall i \in [0;k]$ with probability at least $(1-\delta)^{n}$, and $x_k' \in X \setminus X_S$, i.e., 
$ \mathbb{P}\{(x_i'', \hat x_i'')\in R_x, \forall i \in [0;k]\} \geq (1-\delta)^{k}$.

Let us consider again the 
random trajectory $\xi_{x_0\nu}$ in $\Sigma$ and its corresponding output trajectory $\zeta_{x_0\nu}$, and define the events: 
$\mathcal E_1:=\{(x_i', \hat x_i'')\in R_x, \forall i \in [0;n]\}$; $\mathcal E_2:=\{\zeta_{x_0\nu} \in  \hat{\mathcal{H}} ^{\epsilon}_{\mathcal{C}_\varepsilon} \}$; and $\mathcal E_3:=\{\zeta_{x_0\nu} \in \mathcal{H}_{\mathcal{C}_{(\varepsilon+2\epsilon)}}^n  \}$. 
Note that $\mathcal E_1$ further implies that 
\begin{align}\label{ieq2cur} 
    \max_{i \in [0;n]} \Vert h(x_i') - \hat h(\hat x_i'') \Vert \leq \epsilon,
\end{align} 
by condition 2 in Definition~\ref{CurRinter}. 
Now consider $\zeta_{x_0\nu} =(y_0,\dots, y_n)$ with $\mathcal E_2:=\{\zeta_{x_0\nu} \in  \hat{\mathcal{H}} ^{\epsilon}_{\mathcal{C}_\varepsilon} \}$ holds,
by leveraging the above implications from $\mathcal E_2$, we get that
there exists $(x_0', \dots, x_n')$ in $\Sigma$ being current-state secure with none of its prefix revealing current-state secret as in \eqref{ieq1cur}, such that 
$\max_{i \in [0;n]} \Vert h(x_i') - y_i \Vert \leq \varepsilon+2\epsilon$,
where the inequality is obtained by combining  \eqref{abssecureoutput2cur}, \eqref{ieq1cur}, and \eqref{ieq2cur}.
This implies that $\zeta_{x_0\nu} \in \mathcal{H}_{\mathcal{C}_{(\varepsilon+2\epsilon)}}^n$ holds.
Therefore, we obtain that 
$\mathcal E_1 \cap \mathcal E_2 \Longrightarrow \mathcal E_3$ holds, and thus $\mathbb{P}(\mathcal E_1 \cap \mathcal E_2) \leq \mathbb{P}(\mathcal E_3)$. Therefore, we get $\mathbb{P}(\zeta_{x_0\nu} \in 	\mathcal{H}_{\mathcal{C}_{(\varepsilon+2\epsilon)}}^n) \stackrel{\eqref{Prob_ep}}{\geq} (1-\delta)^{n}(\lambda- \gamma_\delta) = (1-\gamma_\delta)(\lambda- \gamma_\delta)$.
By leveraging again \eqref{eqp2}, we have $\mathbb{P}(\xi_{x_0\nu} \in L_{\mathcal{C}_{(\varepsilon+2\epsilon)}}^n) \geq (1-\gamma_\delta)(\lambda- \gamma_\delta)$. Thus, we can conclude that $\Sigma$ is  $(\varepsilon+2\epsilon,\lambda_\delta)$-approximate current-state opaque within time horizon $n \in \mathbb{N}$, with $\lambda_\delta = (1-\gamma_\delta)(\lambda- \gamma_\delta)$.
$\hfill\blacksquare$


{\bf Proof of Theorem~\ref{mainthem}.}
We start with showing condition 1 in Definition \ref{InitialRinter}. Consider any secret initial state $x_{0} \in X_{0} \cap X_{S}$ in $\Sigma$,
	since $\eta\leq span( X_S)$, there always exists a representative point $\hat x_{0} = {\it\Pi}_x(x_{0})$ in $\hat{\Sigma}$ inside the set $\hat X_{0} \cap \hat X_{S}$ by the  construction of sets  $\hat{X}_0 = [ X]_{\eta}$ and $\hat{X}_S = [ X_S^{\theta}]_{\eta}$. 
 Note that $\Vert x_{0}-\hat x_{0}\Vert \leq \eta$ holds by \eqref{eq:Pi_delta}. By combining \eqref{Con555a} and \eqref{quanti}, we have $ 
		V(x_{0},\hat x_{0})\leq \overline{\alpha} (\Vert x_{0}-\hat x_{0}\Vert ) \leq \overline{\alpha} (\eta) \leq \underline{\alpha} \circ {\ell^{-1}}(\varepsilon)$,
	which implies $(x_{0},\hat x_{0})\in R_{x}$ by the definition of $R_{x}$ in \eqref{localsr}. Hence, condition 1(a) is satisfied readily.
	For condition 1(b), for every $\hat x_{0} \in \hat X_{0} \setminus \hat X_{S}$, by choosing $x_{0} = \hat x_{0}$ which is also inside $X_{0} \setminus X_{S}$, we again get $V(x_{0},\hat x_{0}) = 0 \leq \underline{\alpha} \circ {\ell^{-1}}(\varepsilon)$ which implies that $(x_{0},\hat x_{0}) \in R_{x} $ holds.
	We proceed with showing condition 2. 
	Combining the Lipschitz assumption  \eqref{lipschitz} and \eqref{Con555a}, one gets for any $(x,\hat x) \in R_x$,
 $ \Vert h(x)-h(\hat x ) \Vert \stackrel{\eqref{lipschitz}}{\leq} \ell(\Vert x-\hat x\Vert) \stackrel{\eqref{Con555a}}{\leq} \ell\circ \underline \alpha^{-1}(V(x,\hat x)) \stackrel{\eqref{localsr}}{\leq} \epsilon$,
 which shows that condition 2 is satisfied. 
 
Next, we show condition 3 also holds.
By taking the conditional expectation from \eqref{Eq65}, for any $(x , \hat x) \in R_x$, for $u \in U$, there exists $\hat u \in \hat U$, and vice versa, such that 
$\mathbb{E}\Big[V(f(x,u,\varsigma),\hat f(\hat x, \hat u,\varsigma))\,\big|\,x,\hat x,u,\hat u\Big] 
 - \mathbb{E}\Big[V(f(x,u,\varsigma), f(\hat x, \hat u,\varsigma))\,\big|\,x,\hat x,u,\hat u\Big] 
 \leq \mathbb{E}\Big[\gamma (\Vert \hat f(\hat x, \hat u,\varsigma)-f(\hat x, \hat u,\varsigma)\Vert)\,\big|\,x,\hat x,u,\hat u\Big]$,  
where $\hat f(\hat x, \hat u, \varsigma) =  {\it {\Pi}}_x(f(\hat x, \hat u, \varsigma))$. Using \eqref{eq:Pi_delta}, we further get 
\begin{align} \notag
&\mathbb{E}\Big[V(f(x,u,\varsigma),\hat f(\hat x, \hat u,\varsigma))\,\big|\,x,\hat x,u,\hat u\Big]\\ \label{ineq2}
&- \mathbb{E}\Big[V(f(x,u,\varsigma), f(\hat x, \hat u,\varsigma))\,\big|\,x,\hat x,u,\hat u\Big] \leq \gamma(\eta).
\end{align}
By \eqref{Con555b}, we have {\small $ 
\mathbb{E}\Big[V(f(x,u,\varsigma), f(\hat x, \hat u,\varsigma))\,\big|\,x,\hat x,u,\hat u\Big]$  $\leq (\mathcal{I}_d-\bar{\kappa})(V(x,\hat x))+\bar \rho(\Vert u-u'\Vert)$}. 
By further using \eqref{ineq2} and \eqref{eq:Pi_delta}, it follows that 
\begin{align} \notag
&\mathbb{E}\Big[V(f(x,u,\varsigma),\hat f(\hat x, \hat u,\varsigma))\,\big|\,x,\hat x,u,\hat u\Big]\\ \label{ineq4}
&\leq  (\mathcal{I}_d-\bar{\kappa})(V(x,\hat x))+\bar \rho(\mu) +  \gamma(\eta).
\end{align}
Applying Markov's inequality \cite{durrett2019probability} to \eqref{ineq4}, we get the following inequality chain 
	\begin{align} \notag 
&\mathbb{P}\{ V(f(x,u,\varsigma),\hat f(\hat x, \hat u,\varsigma)) \leq   \underline{\alpha} \circ {\ell}^{-1}(\epsilon) \,\big|\,x,\hat x,u,\hat u\} \\ \notag
		&= 1- \mathbb{P}\{ V(f(x,u,\varsigma),\hat f(\hat x, \hat u,\varsigma))\geq   \underline{\alpha} \circ {\ell}^{-1}(\epsilon) \,\big|\,x,\hat x,u,\hat u\} \\ \notag
		&\geq 1 - \frac{1}{\underline{\alpha} \circ {\ell}^{-1}(\epsilon)}\mathbb{E}\Big[V(f(x,u,\varsigma),\hat f(\hat x, \hat u,\varsigma))\,\big|\,x,\hat x,u,\hat u\Big] \\ \notag
		& \stackrel{\eqref{localsr}}{\geq} 1 - \frac{1}{\underline{\alpha} \circ {\ell}^{-1}(\epsilon)} \Big [(\mathcal{I}_d-\bar{\kappa}) \circ \underline{\alpha} \circ {\ell}^{-1}(\epsilon)+\bar \rho(\mu) +\gamma(\eta) \Big] \\ \label{markovineq}
	&\stackrel{\eqref{quanti}}{\geq} 1-\delta. 
	\end{align}
	Note that the last inequality in \eqref{markovineq} is achieved from \eqref{quanti}, where one has
$\frac{1}{\underline{\alpha} \circ {\ell}^{-1}(\epsilon)} \{(\mathcal{I}_d-\bar{\kappa})\circ \underline{\alpha} \circ {\ell}^{-1}(\epsilon)+\bar \rho(\mu)
		+\gamma(\eta) \} \leq \delta$.
Hence, condition 3 of Definition \ref{InitialRinter} is also satisfied, which completes the proof.
$\hfill\blacksquare$

{\bf Proof of Theorem~\ref{mainthem_cur}.}
We start with showing condition 1 in Definition \ref{CurRinter}. Consider any initial state $x_{0} \in X_{0} $ in $\Sigma$,
	since $\eta\leq span(X_0)$, there always exists a representative point $\hat x_{0} = {\it\Pi}_x(x_{0})$ in $\hat{\Sigma}$ by the construction of $\hat{X}_{0}$ such that $\Vert x_{0}-\hat x_{0}\Vert \leq \eta$ holds by \eqref{eq:Pi_delta}. By \eqref{Con555a} and \eqref{quanticur}, we have  
		$V(x_{0},\hat x_{0})\leq \overline{\alpha} (\Vert x_{0}-\hat x_{0}\Vert ) \leq \overline{\alpha} (\eta) \leq \underline{\alpha} \circ {\ell^{-1}}(\epsilon)$,
	which implies $(x_{0},\hat x_{0})\in R_{x}$ by the definition of $R_{x}$ in \eqref{localsr_cur}. 
 	For condition 1(b), for every $\hat x_{0} \in \hat X_{0} $, by choosing $x_{0} = \hat x_{0}$ which is also inside $X_{0}$, we again get $V(x_{0},\hat x_{0}) = 0 \leq \underline{\alpha} \circ {\ell^{-1}}(\varepsilon)$ which implies that $(x_{0},\hat x_{0}) \in R_{x} $ holds. 
 Hence, condition 1 is satisfied readily.
 Conditions 2, 3(a) and 3(c) can be shown following the same reasoning as in Theorem~\ref{mainthem}, thus is omitted here. Finally, let us show that conditions 3(b) and 3(d) in Definition \ref{CurRinter} hold as well. 
 Consider any pair of states $(x,\hat x) \in R_{x}$, and suppose that under an input $u \in U$, we get $x^+ = f(x,u, \varsigma) \in X_S$.   
 Similar to the proof of condition 3(a), we can show that by choosing $\hat u = u$, we have $(x^+, \hat x^+) \in R_x$ with probability greater than $1-\delta$, where $\hat x^+ = \hat f(\hat x,\hat u, \varsigma)$. 
Since $(x^+, \hat x^+) \in R_x$, by combining \eqref{localsr_cur} with \eqref{Con555a} and \eqref{quanticur2}, we have $\Vert x^+ -\hat x^+\Vert \leq \underline{\alpha}^{-1}(V(x^+, \hat x^+)) \leq {\ell^{-1}}(\epsilon)\leq \theta$. 
  Moreover, by the construction of the secret set in the finite abstraction, one has $\hat X_{S} = [X_S^{\theta}]_{\eta}$ with the inflation parameter $\theta$  satisfying ${\ell^{-1}}(\epsilon) \leq \theta$ and $0 \!<\! \eta \!\leq\! \text{min} \{span(X_S),span({X} \setminus X_S)\}$.  
 This further implies that $\hat x^+ \in \hat{X}_S$ holds as well. Thus, condition 3(b) is satisfied. Condition 3(d) can be proved in a similar way.  Therefore, we conlude that condition 3 of Definition \ref{CurRinter} is satisfied, which completes the proof.
$\hfill\blacksquare$

\vspace{-0.3cm}

\bibliographystyle{IEEEtran}
\bibliography{biblio}

\end{document}